\documentclass[
showkeys,12pt,
preprint,preprintnumbers,nofootinbib,
groupedaddress,superscriptaddress,amsmath,amssymb]{revtex4}
\usepackage{graphicx}
\usepackage{dcolumn}
\usepackage{bm}
\usepackage{amssymb}
\usepackage{amsmath}
\usepackage{epsfig}    
\usepackage{color}
\usepackage{slashed}
\usepackage{hhline}
\usepackage{comment}
\usepackage{braket}

\def\be{\begin{equation}}
\def\ee{\end{equation}}
\newcommand{\bea}{\begin{eqnarray}}
\newcommand{\eea}{\end{eqnarray}}
\newcommand{\nn}{\nonumber}

\numberwithin{equation}{section}

\begin{document}
\allowdisplaybreaks[2]

\title{
{Three loop neutrino model with isolated $k^{\pm\pm}$}
}
\preprint{KIAS-P15035
}

\author{Kenji Nishiwaki}
\email{nishiken@kias.re.kr}
\affiliation{School of Physics, Korea Institute for Advanced Study,\\ Seoul 130-722, Republic of Korea}

\author{Hiroshi Okada}
\email{macokada3hiroshi@gmail.com}
\affiliation{School of Physics, Korea Institute for Advanced Study,\\ Seoul 130-722, Republic of Korea}
\affiliation{Physics Division, National Center for Theoretical Sciences, National Tsing-Hua University, Hsinchu 30013, Taiwan}

\author{Yuta Orikasa}
\email{orikasa@kias.re.kr}
\affiliation{School of Physics, Korea Institute for Advanced Study,\\ Seoul 130-722, Republic of Korea}
\affiliation{Department of Physics and Astronomy, Seoul National University, Seoul 151-742, Republic of Korea}

\date{\today}

\begin{abstract}
We propose a {three loop} radiative neutrino mass scenario with an {\it isolated} {doubly charged} singlet scalar $k^{\pm\pm}$ without couplings to the charged leptons, while {two other singly charged} scalars $h_1^\pm$ and $h_2^\pm$ attach to them.
In this setup, the lepton flavor violation originating from $k^{\pm\pm}$ exchanges is {suppressed} and the model is less constrained, where some couplings can take sizable values.
As reported in our previous work~\cite{Hatanaka:2014tba}, the loop suppression factor at the three loop level would be too strong and realized neutrino masses in a {three loop} scenario could be smaller than the observed minuscule values.
The sizable couplings can help us to {\it enhance} neutrino masses without drastically large scalar trilinear couplings appearing {in} a neutrino mass matrix, which tends to drive the vacuum stability {to become} jeopardized at the {one loop} level.
Now the {doubly charged} scalar $k^{\pm\pm}$ has {less} constraint via lepton flavor violation and the vacuum can be quite stable, and thus a few hundred GeV mass in $k^{\pm\pm}$ is possible, which is within the LHC reach and this model can be tested in the near future.
Note that the {other} $h_1^\pm$ and $h_2^\pm$ should be heavy at least around a few TeV.
We suitably arrange the charges of an additional global $U(1)$ symmetry, where the decay constant of the associated Nambu-Goldstone boson can be around a TeV scale consistently.
Also, this model is indirectly limited through a global analysis on results of the LHC Higgs search and issues on a dark matter candidate, the lightest Majorana neutrino.
After $h_1^\pm$ and $h_2^\pm$ {are decoupled}, this particle couples to the {standard model} particles only through two {charge parity even} scalars in theory and thus information on this scalar sector is important.
Consistent solutions are found, but a part of them is now on the edge.
\end{abstract}
\maketitle


\newpage

\section{Introduction}

Recently, the second round of the physics run at the Large Hadron Collider~(LHC) started and a magnificent operation for exploring the scalar sector describing the {electroweak}~(EW) scale was launched.
The greatest achievement at the first run of {the} LHC is the observation of the new Higgs-like scalar boson, which was the last missing piece of the {standard model}~(SM), around $125\,\text{GeV}$ by the ATLAS and the CMS experiments~\cite{Aad:2012tfa,Chatrchyan:2012ufa}.

The Yukawa couplings to heavy fermions of {the} SM, namely the top {and} bottom quarks and the tau {lepton} have been surveyed with good precision, whereas the lighter states are still mysterious from {both} the experimental and theoretical point of view.
The extremely tiny observed masses of active components of the neutrinos would be {key to} investigating the scalar sector theoretically because we should accept at least $10^{12}$-order {hierarchy in the Higgs Yukawa couplings} for describing the neutrino nature within {the} SM.

One of the most stimulating ideas for {addressing} this issue is radiative generations of the neutrino profiles.
Loop suppression factors should appear in the neutrino masses in this type of {scenario}, which help to alleviate the {hierarchy in} couplings.
Another motivation for this direction is that the continuous and/or discrete symmetries ensuring the loop origin could also guarantee the existence of a (or multiple) dark matter~(DM) candidate(s).
Following the landmarks~\cite{Zee,Li,zee-babu,Krauss:2002px,Ma:2006km}, {recently}
a variety of works on radiative seesaw model have been done~\cite{Hambye:2006zn}--\cite{Kashiwase:2015pra}, where we can also find
studies {emphasizing} non-Abelian discrete symmetry~{\cite{Ahn:2012cg,Ma:2012ez,Kajiyama:2013lja,Hernandez:2013dta,Ma:2014eka,Aoki:2014cja,Ma:2014yka,Ma:2015pma}},
radiative generation of quark/charged lepton masses~\cite{Hernandez:2013hea,Ma:2013mga,radlepton1,radlepton2,Okada:2014nsa},
operator analysis~\cite{Bonnet:2012kz,Sierra:2014rxa},
radiative models accompanying conformal EW symmetry breaking~\cite{Davoudiasl:2014pya,Lindner:2014oea,Okada:2014nea},
and others~\cite{Babu:2002uu,AristizabalSierra:2006ri,AristizabalSierra:2006gb,Nebot:2007bc,Bouchand:2012dx,Kajiyama:2013sza,McDonald:2013hsa,Ma:2014cfa,Schmidt:2014zoa,Herrero-Garcia:2014hfa,Ahriche:2014xra,Long1,Long2}.

From the naturalness point of view, higher-loop generation is better.
The first {three loop} model for a natural explanation of the neutrino profiles was proposed in Ref.~\cite{Krauss:2002px} and the following works continue~\cite{Aoki:2008av,Gustafsson,Kajiyama:2013lja,Ahriche:2014xra,Ahriche:2014cda,Ahriche:2014oda,Chen:2014ska,Okada:2014oda,Hatanaka:2014tba,Jin:2015cla,Culjak:2015qja,Geng:2015coa,Ahriche:2015wha}, where in such situations, couplings related with the neutrino masses can be close to unity compared with those in models with {one or two loop} level generation.

On the other hand, {three loop} generations could face problems owing to {the} largeness of couplings.
As discussed in Ref.~\cite{Hatanaka:2014tba}, the {three loop} suppression factor $1/(4\pi)^6$ sometimes looks very strong and we probably {\it enhance} {part of} couplings for {a suitable} realization of the neutrino masses.
Considering large {Yukawa-type} couplings {with lepton flavor violation~(LFV)} {tend} to result in unacceptable enhancements in {LFV} processes.
Hence, choosing 
sizable scalar trilinear couplings appearing in the neutrino mass matrix, {which do not generate LFV directory,} seems to be a reasonable prescription.
But this diagnosis could {still} be at least partly a misjudgment.
When we go for {one loop} level, as shown in Ref.~\cite{Hatanaka:2014tba}, these substantial trilinear couplings give drastic negative contributions to quartic couplings of charged scalar bosons and the vacuum can be {threatened} with being destabilized.
The last option would be to accept decouplings of additional charged bosons, namely around $10\,\text{TeV}$, where the neutrino profile itself can be suitably generated.
But finding 
clear signals at colliders becomes very difficult even at the updated {$13$} or $14\,\text{TeV}$ LHC.

To circumvent the situation, we propose a refined {three loop} model by use of a global $U(1)$ symmetry without additional discrete symmetry.
A key point is introducing additional Majorana neutrinos, which can violate lepton {flavors} in the fermion line inside diagrams describing the neutrino masses.
It is important that with the above setup, a {doubly charged} scalar $k^{\pm\pm}$ no longer needs to have direct couplings to the charged leptons.
We mention 
that in our previous model without Majorana neutrinos in Ref.~\cite{Hatanaka:2014tba}, like in the Zee-Babu model~\cite{zee-babu}, a {doubly charged} scalar should attach the charged leptons to generate violation in lepton {flavors}, where tree-level lepton flavor violating processes are generated and we cannot put large values {in the corresponding couplings consistently.}

In the present $k^{\pm\pm}$-{\it isolated} scenario, the {doubly charged} scalar $k^{\pm\pm}$ is quarantined from the charged lepton sector at the leading order by a suitable choice of global $U(1)$ charges.
Now, $k^{\pm\pm}$ cannot contribute to phenomena with {LFV} at the leading order and constraints on couplings are weakened.
Consequently, the scalar trilinear couplings can take smaller values and the vacuum stability would not be so serious even when we consider a few hundred GeV $k^{\pm\pm}$, the detection of which can be an evident signal for probing this scenario at the LHC {experiments}.

This paper is organized as follows.
In {Sec.}~\ref{sec:mainbody}, first we introduce our setups and subsequently, we discuss miscellaneous issues in this model, namely, forms of scalar masses and mixings, properties of the Nambu-Goldstone~(NG) boson associated with the breakdown of the global $U(1)$ symmetry, sizable correction via charged scalars to vacuum stability, form of the active neutrino mass matrix at the {three loop level, and} details on processes accompanying LFV in order.
In the following {Sec.}~\ref{sec:neutrino_fitting}, after having a discussion on analogies with the Zee-Babu model~\cite{zee-babu}, we execute parameter scans both in the normal and inverted hierarchies.
In {Sec.}~\ref{sec:LHCglobalfit}, we make global fits of signal strengths of the Higgs production in various channels announced by the ATLAS and the CMS experiments, which restrict possible values of the mass of the {doubly charged} scalar and the mixing angle between the SM {Higgs} boson and an additional {charge parity ($CP$)} even scalars, whose vacuum expectation value~(VEV) breaks the global $U(1)$ symmetry.
In {Sec.}~\ref{sec:DM calculation}, we discuss properties of the dark matter candidate of this scenario, {which is} the lightest {right-hand} neutrino{,} through a relic density calculation, an estimation of {the} invisible decay width of the observed $125\,\text{GeV}$ scalar and an evaluation of spin-independent direct detection cross section.
Section~\ref{sec:conclusion} is devoted to summarizing results and making conclusions.
{In Appendix~\ref{sec:appendix_loopfunction}, we give analytic forms of the loop functions describing lepton-flavor-violating processes.
In Appendix~\ref{sec:appendix_width}, a part of partial decay widths of the two {$CP$} even scalars with nontrivial forms {is} described.
In Appendix~\ref{sec:appendix_amplitude}, we summarize the averaged matrix elements squared for relic density calculation.
}

\section{Basic issues on the scenario \label{sec:mainbody}}

\subsection{Model setup}

We discuss a {three loop} induced radiative neutrino model with {a} $U(1)$ global symmetry. 
The particle contents and their charges are shown in {Table}~\ref{tab:1}. 
We introduce three Majorana fermions $N_{R_{1,2,3}}$ and new bosons; one gauge-singlet neutral boson $\Sigma_0$, two
{singly charged} singlet scalars ($h^\pm_1, h^\pm_2$), and one gauge-singlet {doubly charged} boson $k^{\pm\pm}$ to {the} SM.
We assume that  only the SM-like Higgs $\Phi$ and the additional neutral scalar $\Sigma_0$ have
VEVs, which are symbolized as $\langle\Phi\rangle\equiv v/\sqrt2$ and $\langle\Sigma_0\rangle\equiv v'/\sqrt2$, respectively.
{Here, we set $v$ as $\simeq 246\,\text{GeV}$.
It is natural that $v'$ is greater than $v$ to some extent to evade a large mixing, which is constrained by the LHC data{; see Sec.}~\ref{sec:LHCglobalfit} for details.}
$x\,(\neq0)$ is an arbitrary number of the charge of the hidden $U(1)$ symmetry,\footnote{Here, we assume that  this $U(1)$ symmetry is global. {However,} one can straightforwardly move to the local one by introducing {additional fermions}}
and under the assignments, neutrino mass matrix is generated at the {three loop} level.
{Note that a remnant $Z_2$ symmetry remains after the hidden $U(1)$ symmetry breaking and
the particles $N_{R_{1,2,3}}$ and $h^\pm_2$ have negative parities.
Then, when a Majorana neutrino is the lightest among them, the stability is accidentally ensured.}

\begin{widetext}
\begin{center} 
\begin{table}[tbc]
\begin{tabular}{|c||c|c|c||c|c|c|c|c|}\hline\hline  
&\multicolumn{3}{c||}{Lepton Fields} & \multicolumn{5}{c|}{Scalar Fields} \\\hline
{Characters} & ~$L_{L_i}$~ & ~$e_{R_i}$~ &~$N_{R_i}$~ & ~$\Phi$~ &
 ~$\Sigma_0$~ & ~$h^+_1$~  & ~$h^{+}_2$~ & ~$k^{++}$~ \\\hline 
$SU(3)_C$ & $\bm{1}$ & $\bm{1}$ & $\bm{1}$ & $\bm{1}$ & $\bm{1}$ & $\bm{1}$ & $\bm{1}$ & $\bm{1}$ \\ \hline
$SU(2)_L$ & $\bm{2}$ & $\bm{1}$&  $\bm{1}$&$\bm{2}$ & $\bm{1}$ &$\bm{1}$  &$\bm{1}$  &$\bm{1}$ \\\hline 
$U(1)_Y$ & $-1/2$ & $-1$ & $0$ & $1/2$  & $0$  & $1$  & $1$ & $2$ \\\hline
{$U(1)$} & $0$ & $0$ & $-x$ & $0$  & $2x$   & $0$ & $x$ & $2x$  \\\hline
\end{tabular}
\caption{Contents of lepton and scalar fields
and their charge assignment under $SU(3)_C \times SU(2)_L\times U(1)_Y\times {U(1)}$, {where $U(1)$  is an additional global symmetry and} $x\neq 0$.
The subscripts found in the lepton fields $i \,(=1,2,3)$ indicate generations of the fields.
{The bold letters emphasize that these numbers correspond to representations of the Lie groups of the NonAbelian gauge interactions.}
}
\label{tab:1}
\end{table}
\end{center}
\end{widetext}

The relevant Lagrangian for Yukawa sector $\mathcal{L}_{Y}$ and scalar potential $\mathcal{V}$ allowed under the global symmetry {is} given as
\begin{align}
{-} \mathcal{L}_{Y}
&=
(y_\ell)_{ij} \bar L_{L_i} \Phi e_{R_j}  + \frac12 (y_{L})_{ij} \bar L^c_{L_i} L_{L_j} h^+_1  + (y_{R})_{ij} \bar N_{R_i} e^c_{R_j} h_2^{-}   +  \frac12 (y_N)_{ij} \Sigma_0 \bar N_{R_i}^c N_{R_j} 
+\rm{h.c.} {,} \\ 
\mathcal{V}
&= 
 m_\Phi^2 |\Phi|^2 + m_{\Sigma}^2 |\Sigma_0|^2 + m_{h_1}^2 |h^+_1|^2  + m_{h_2}^2 |h_2^{+}|^2   + m_{k}^2 |k^{++}|^2 
 \nn\\
&+ \Bigl[
 \lambda_{11}  \Sigma_0^* h^-_1 h^-_1 k^{++} +  \mu_{22} h^+_2 h^+_2 k^{--}   + {\rm h.c.}
 \Bigr]
  +\lambda_\Phi |\Phi|^{4} 
  + \lambda_{\Phi\Sigma} |\Phi|^2|\Sigma_0|^2 
 +\lambda_{\Phi h_1}  |\Phi|^2|h^+_1|^2  
 \nn\\
& +\lambda_{\Phi h_2}  |\Phi|^2|h^+_2|^2 +\lambda_{\Phi k}  |\Phi|^2|k^{++}|^2  
  + \lambda_{\Sigma} |\Sigma_0|^{4} +
 \lambda_{\Sigma h_1}  |\Sigma_0|^2|h^+_1|^2  
  +\lambda_{\Sigma h_2}  |\Sigma_0|^2|h^+_2|^2 
   \nn\\&
 +\lambda_{\Sigma k}  |\Sigma_0|^2|k^{++}|^2  
  + \lambda_{h_1} |h_1^{+}|^{4}
  {+} \lambda_{h_1 h_2}  |h_1^{+}|^2|h^+_2|^2 +\lambda_{h_1 k}  |h_1^{+}|^2|k^{++}|^2  
 \nn\\
&+
 \lambda_{h_2} |h_2^{+}|^{4} + \lambda_{h_2 k}  |h_2|^2|k^{++}|^2  
+ \lambda_{k} |k^{++}|^{4} 
,
\label{HP}
\end{align}
where the indices $i,j$ indicate matter generations {and} the superscript $``c"$ means charge conjugation.\footnote{For $SU(2)_L$ doublets, charge conjugation is defined with the $SU(2)_L$ rotation described by a Pauli matrix as $i\sigma_2$.}
$y_L$, $y_R$ and $y_N$ are antisymmetric, general, symmetric three-by-three matrices, respectively.
The first term of $\mathcal{L}_{Y}$ generates the charged-lepton masses {following} the SM manner.
Majorana mass terms are derived from the fourth one after $\Sigma_0$ obtains {a} VEV.
{Note that this VEV also {generates} an effective trilinear coupling $\mu_{11} \left({\text{in front of }} h_1^- h_1^- k^{++}\right)$  from the quartic coupling $\lambda_{11} \left({\text{in front of }} \Sigma_0^\ast h_1^- h_1^- k^{++} \right)$, where {the} coefficient is given as}
\begin{align}
\mu_{11} = \lambda_{11} \langle \Sigma_0^\ast \rangle = \lambda_{11} {\frac{v'}{\sqrt{2}},}
	\label{form_mu11}
\end{align}
{where we use the {parametrization} of $\langle \Sigma_0 \rangle$ declared in Eq.~(\ref{component}) in the next subsection.}
We assume the following two things: (i) $\lambda_{11}$ and $\mu_{22}$ are positive real; (ii) $y_N$ is diagonal and obeys the hierarchy $(y_N)_{11} < (y_N)_{22} < (y_N)_{33}$ among positive-real parameters, which means that a generated Majorana mass matrix for $N_{R}$ is also diagonal one and the mass ordering is $M_{N_{1}} < M_{N_{2}} < M_{N_{3}}$.
The concrete forms of the masses are
\begin{align}
M_{N_{1}} = \frac{v'}{\sqrt{2}} (y_N)_{11},\quad
M_{N_{2}} = \frac{v'}{\sqrt{2}} (y_N)_{22},\quad
M_{N_{3}} = \frac{v'}{\sqrt{2}} (y_N)_{33}.
	\label{Majonara_mass_form}
\end{align}

\subsection{Mass eigenvalues and eigenstates of scalars}

The neutral scalar fields are {parametrized} as 
\begin{align}
&\Phi =\left[
\begin{array}{c}
w^+\\
\frac{v+\phi+i z}{\sqrt2}
\end{array}\right],\quad 
\ 
\Sigma_0=\frac{v'+\sigma}{\sqrt{2}}e^{iG/v'}
.   
\label{component}
\end{align}
where $v \simeq 246\,\text{GeV}$ is {the} VEV of the Higgs doublet field, and $w^\pm$
and $z$ are (would-be) NG {bosons}
{that} are absorbed as the longitudinal components of {the} $W$ and $Z$ bosons, respectively.
Requiring the tadpole conditions, $\partial\mathcal{V}/\partial\phi|_{v}=0$ and $\partial\mathcal{V}/\partial\sigma|_{v'}=0$,
{the} {resultant} mass matrix squared of the {$CP$} even components $(\phi,\sigma)$ 
 is given by
\begin{equation}
m^{2} (\phi,\sigma) = \left[%
\begin{array}{cc}
  2\lambda_\Phi v^2 & \lambda_{\Phi \Sigma}vv' \\
  \lambda_{\Phi \Sigma}vv' & 2\lambda_{\Sigma}v'^2 \\
\end{array}%
\right] = \left[\begin{array}{cc} \cos\alpha & \sin\alpha \\ -\sin\alpha & \cos\alpha \end{array}\right]
\left[\begin{array}{cc} m^2_{h} & 0 \\ 0 & m^2_{H}  \end{array}\right]
\left[\begin{array}{cc} \cos\alpha & -\sin\alpha \\ \sin\alpha &
      \cos\alpha \end{array}\right],
      \label{eq:CP-even_matrix}
\end{equation}
where $h$ is the SM-like Higgs {($m_h = 125\,\text{GeV}$)} and $H$ is an additional {$CP$ even} Higgs mass
eigenstate. The mixing angle $\alpha$ is {determined as} 
\be
\sin 2\alpha=\frac{2\lambda_{\Phi \Sigma} v v'}{{m^2_H-m_h^2}}.
\label{eq:CP-even_mixing}
\ee
The neutral bosons $\phi$ and $\sigma$ are rewritten in terms of the mass eigenstates $h$ and $H$ as
\begin{eqnarray}
\phi = h\cos\alpha + H\sin\alpha,
\quad
\sigma =- h\sin\alpha + H\cos\alpha.
\label{eq:mass_weak}
\end{eqnarray}
{A} NG boson $G$ emerges due to the spontaneous symmetry breaking of
the global $U(1)$ symmetry. 
The mass {eigenvalues} for the {singly charged} bosons $h_1^\pm$, $h_2^\pm$ and the {doubly charged boson} $k^{\pm\pm}$ are given as
\begin{align}
&m^{2}_{h^{\pm}_1} = m_{h_1}^{2}  + \frac12 (\lambda_{\Phi h_1} v^{2}+\lambda_{\Sigma h_1} v'^{2}), \quad 
m^{2}_{h^{\pm}_2} = m_{h_2}^{2}  + \frac12 (\lambda_{\Phi h_2} v^{2}+\lambda_{\Sigma h_2} v'^{2}), \notag \\ 
&m^{2}_{k^{\pm\pm}} = m_{k}^{2}  + \frac12 (\lambda_{\Phi k} v^{2}+\lambda_{\Sigma k} v'^{2}),
	\label{eq:mass_eigenvalues}
\end{align}
where the three charged particles are not mixed due to the {symmetries} of the system and thus they themselves are mass eigenstates.

\subsection{Issues on the {Goldstone} boson}

Accompanying the spontaneous breakdown of a $U(1)$ global symmetry, {a} NG boson emerges as an {almost} massless state in theory, which could play {significant} roles in particle physics and cosmology~\cite{Weinberg:2013kea}.
Like the usual Majoron case~\cite{majoron}, our NG boson $G$ communicates with the lepton sector.

An important characteristic of $G$ is that, as described in {Table}~\ref{tab:1}, the lepton doublets and the charged lepton singlets do not hold nonzero charges of the global $U(1)$.
This means that no anomaly-induced interaction to two photons is generated in our setup, which puts a significant constraint on the decay constant of NG bosons~\cite{Cadamuro:2011fd}.
Thus, in the present scenario, we can choose ``lower" values around a TeV scale {without doing any} harm.

Another route for constraining models via NG bosons is through the active-sterile component mixing as through the lepton-flavor-violating transition like  $\mu^{-} \to e^{-} \, \text{(NG)}$ seen in Majoron {seesaw} scenarios, {\it e.g.}, discussed in Ref.~\cite{Pilaftsis:1993af}.
Different from such a situation in our case, the active and the sterile components cannot mix {with}
each other since this mixture is prohibited by the residual $Z_2$ symmetry after the global $U(1)$ breaking shown in {Table}~\ref{tab:1}.
Then {the} absence of this type of {constraint} is assured via the accidentally remaining symmetry.
The neutrinoless double beta decay {via $W$ exchanges} does not restrict our scenario since {the} sequence {with $W$ boson} also requires the active-sterile mixing.
{Note that the three additional charged singlet scalars have no direct coupling to the quarks and {are} therefore ineffective.}

In contrast, the NG boson $G$ couples to the corresponding {$CP$} even component $\sigma$, which should {mix} 
with the {Higgs} component of the doublet $\Phi$.
This means that $G$ can contribute to physics associated with the {$CP$} even scalars.
As we see in {Sec.}~\ref{sec:DM calculation}, the pair annihilation process of the dark matter candidate $N_{R_1}$ is just an example.\footnote{Another interesting topics is collider searches for a NG boson through invisible channels (subsequent decays from {$CP$} even scalars)~\cite{Cheung:2013oya}.}

Finally, we briefly comment on possible bounds from cosmological issues.
For example, an effect on cosmic microwave background via cosmic string generated by the spontaneous breakdown of the global $U(1)$ symmetry possibly put a constraint on our scenario.
The bound discussed in Ref.~\cite{Battye:2010xz} can be {interpreted} as $v' < 10^{15}\,\text{GeV}$, and thus this issue is harmless.
On the other hand, as we discuss {later} in {Sec.}~\ref{sec:scanning}, at least {part} of scalar self couplings tends to be $\mathcal{O}(1)$ (at around the {EW}
scale) owing to the requirements via coexistence of the observed active neutrino profiles and the null observation in {lepton-flavor-violating} currents. 
This trend would lead to blowups of the self couplings {a little bit above the} {lower} scale {that} we focus on in this paper.
{Then} it might not be so fruitful to discuss issues originating from physics at a higher scale.


\subsection{Vacuum stability against charged scalar trilinear couplings}

Vacuum stability has to be especially assured in the Higgs potential against contributions from {electrically charged} bosons ($h_1^{\pm}, h_2^{\pm}, k^{\pm\pm}$). 
However, our model has some loop contributions to leading-order values of these quartic couplings {via the scalar trilinear couplings $\mu_{11}$ and $\mu_{22}$.
When they are large, we should examine the vacuum stability against the effect.}
Here, we examine this issue at the {one loop} level. Let us describe these quartic couplings as {follows,}
\begin{align}
&0 \ {<}\ \lambda_{h_1}=\lambda^{(0)}_{h_1}+\lambda^{(1)}_{h_1},
	\label{eq:vacuumstab_constraint1} \\
&0 \ {<}\  \lambda_{h_2}=\lambda^{(0)}_{h_2}+\lambda^{(1)}_{h_2},
	\label{eq:vacuumstab_constraint2} \\
&0 \ {<}\  \lambda_{k}=\lambda^{(0)}_{k}+\lambda^{(1)}_{k},
	\label{eq:vacuumstab_constraint3}
\end{align}
where the upper indices denote the {numbers} of the order {in loops}, and the one loop contributions are given as
\begin{align}
&\lambda^{(1)}_{h_1}=
-8 |\mu_{11}|^4 F_0(m_{h^{\pm}_1},m_{k^{\pm\pm}}), \\
&\lambda^{(1)}_{h_2}=
-8 |\mu_{22}|^4 F_0(m_{h^{\pm}_2},m_{k^{\pm\pm}}),
\\
&\lambda^{(1)}_{k}=
-4 |\mu_{11}|^4 F_0(m_{h^{\pm}_1},m_{h^{\pm}_1})
-4 |\mu_{22}|^4 F_0(m_{h^{\pm}_2},m_{h^{\pm}_2}),
\end{align}
with
\begin{align}
F_0(m_a,m_b) &= \frac{1}{(4\pi)^2}\int_0^1 dx dy\delta(x+y-1)\frac{x y}{(x m_a^2+y m_b^2)^2} \notag \\
&=
\begin{cases}
\displaystyle \frac{1}{(4\pi)^2} \frac{m_a^2 \left(\log \left(\frac{m_a^2}{m_b^2}\right) -2 \right)+m_b^2
   \left(\log \left(\frac{m_a^2}{m_b^2} \right) +2 \right)}{\left(m_a^2-m_b^2\right){}^3}
   & (\text{for } m_a \not= m_b), \\
\displaystyle \frac{1}{(4\pi)^2} \frac{1}{6m_a^4} & (\text{for } m_a = m_b),
\end{cases}
\label{cdt-vs}
\end{align}
where {the form of $\mu_{11}$ is shown in Eq.~(\ref{form_mu11})} and each of $m_1$ and $m_2$ in $F_0$ represents a mass of propagating fields in the loops.
We {include} these constraints in the numerical analysis later.
To avoid the global minimum accompanying charge breaking, the following  condition should at least be satisfied:
\begin{align}
|\mu_{22}| <   \sqrt{\Lambda}\left[m^2_\Phi + m^2_{h_1} + m^2_{h_2}+ m^2_{k}+ m^2_{\Sigma} \right]^{1/2},
\quad \Lambda\equiv\sum_{i={\rm all\ quartic\ couplings\ including\ \lambda_{11}}}\lambda_i, 
\end{align}
where we assume the simplified configuration, $r\equiv |\Phi|=|h^+_1|=|h^+_2|=|k^{++}|=|\Sigma_0|$ {and the above inequality comes from the requirement that $r$ does not have a finite nonzero value.}
The summation is taken over the coefficients of the $17$ quartic terms in Eq.~(\ref{HP}) including $\Sigma_0^\ast h_1^- h_1^- k^{++}$ and its {Hermitian} conjugate.
When all of these quartic couplings are assumed to take the same value $\lambda$, the above condition is rewritten as
\begin{align}
|\mu_{22}|  \lesssim  4 \sqrt{\lambda} \left[m^2_{h_1} + m^2_{h_2}+ m^2_{k}+ m^2_{\Sigma}   \right]^{1/2},
	\label{form_of_evading_globalminimum}
\end{align}
where the contributions via $m^2_\Phi$ and {$\lambda_\Phi$ are insignificant and thus neglected.}

\subsection{Neutrino mass matrix}

\begin{figure}[t]
\begin{center}
\includegraphics[scale=0.7]{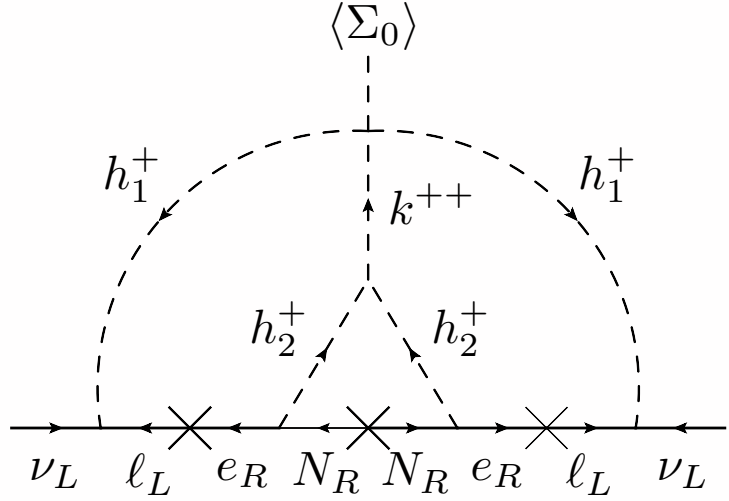}
   \caption{{A schematic description for the radiative generation of neutrino masses.}}
   \label{neutrino-diag}
\end{center}
\end{figure}
A Majorana neutrino mass matrix $m_\nu$ is generated at the {three loop} level via the
diagram shown in Fig.~\ref{neutrino-diag}, which corresponds to the coefficients of the
effective operators, $-\frac{1}{2} (m_{\nu})_{ab} \times \overline{(\nu_{L_a})^c} \nu_{L_b}$.
The form of $(m_{\nu})_{ab}$ is evaluated by a straightforward calculation as
\begin{align}
(m_{\nu})_{ab}
&=
\frac{\mu_{11}\mu_{22}}{(4\pi)^6}
\sum_{i,j,k=1}^{3}
{\frac{1}{M_k^4}}
\left[
 (y_L)_{ai}m_{\ell_i}
 (y^T_{R})_{ik}
 (M_{N_k})
  (y_{R})_{kj}
 m_{\ell_j}
  (y^T_L)_{jb}
 \right] \notag \\
&\quad\quad
\times
F_1\left(
\frac{m_{h^+_1}^2}{{M_k^2}},
\frac{m_{h^+_2}^2}{{M_k^2}},
\frac{m_{\ell_i}^2}{{M_k^2}},
\frac{m_{\ell_j}^2}{{M_k^2}},
\frac{{M_{N_k}}^2}{{M_k^2}},
\frac{m_{k^{\pm\pm}}^2}{{M_k^2}}
\right),
	\label{eq:3loop_neutrino_mass}
\end{align}
where the mass scale ${M_k} = {\rm max}[m_{h^\pm_1},m_{h^\pm_2},m_{\ell_{i}},m_{\ell_{j}},M_{N_k},m_{k^{\pm\pm}}]$ is used for factorizing the loop function $F_1$ as a dimensionless variable.
Here, we take the relationship in masses in our setup, $m_{\ell_i}, m_{\ell_j} < M_{N_k}$, into consideration and then $M_{k}$ has only the index $k$.
$F_1$ is symbolically calculated as follows:
\begin{align} 
&F_1\left(X_1,X_2,X_3,X_4,X_5,X_6\right) 
=
\int \mathbf{d X} \frac{1}{\Delta_1} \frac{1}{(\Delta_2)^2} \frac{\rho}{(\Delta_3)^2}, \\
\int \mathbf{d X} &= \int_0^1 \!\!\! dxdydz \, \delta(x+y+z-1) \int_0^1 \!\!\! d\alpha d\beta d\gamma d\delta \,\delta(\alpha+\beta+\gamma+\delta-1) \int_0^1 \!\!\! d\rho d\sigma d\omega \, \delta(\rho+\sigma+\omega-1),
\end{align}
with
\begin{align}
\Delta_1 &= y(y-1)+z(z-1)+2yz, \\
\Delta_2 &= (\alpha Y+\delta)^2-\delta-\alpha Y^2-\alpha X, \\
\Delta_3 &= \rho A\left(X_1, X_2, X_3, X_5, X_6\right) -\sigma X_4-\omega X_1, \\
A\left(X_1,X_2,X_3,X_5,X_6 \right) &=
-\frac{\alpha((x+y) X_2 + z X_5)}{((\alpha Y+\delta)^2-\delta-\alpha Y^2-\alpha X)(y(y-1)+z(z-1)+2yz)}
\nn\\&
\quad +
\frac{\beta X_1 +\gamma X_3+ \delta X_6}{((\alpha Y+\delta)^2-\delta-\alpha Y^2-\alpha X)},\\
X&=-\left(\frac{y}{y+z}\right)^2+\frac{y(y-1)}{y(y-1)+z(z-1)+2yz},\quad
Y=\frac{y}{y+z}.
\end{align}
Here, note that the shape of $F_1$ is completely the same {as} that in Ref.~\cite{Hatanaka:2014tba} except for the content of $X_5$ in Eq.~(\ref{eq:3loop_neutrino_mass}).

Neutrino mass eigenstates and their mixings are evaluated by reflecting on similarities to the Zee-Babu model~\cite{Herrero-Garcia:2014hfa}.
The structure of the fermion line is similar to the Zee-Babu model~\cite{zee-babu}, that is, a rank two model of the neutrino mass matrix with a massless eigenstate due to the {antisymmetricity} of $y_L$.
Let us describe the neutrino mass matrix as   
\begin{align}
&(m_{\nu})_{ab}=(U_{{\text{PMNS}}} \, m^{\text{diag}}_{\nu} \, U^T_{{\text{PMNS}}})_{ab} \equiv
{\sum_{i,j=1}^3} (y_L)_{ai}\omega_{ij} (y^T_L)_{jb},
	\label{eq:neutrino_massmatrix} \\
& \omega_{ij}=
{\sum_{k=1}^3}\,
m_{\ell_i}
 (y^T_{R})_{ik}
 (\zeta_k M_{N_k})
  (y_{R})_{kj}
 m_{\ell_j}{,}
	\label{eq:definition_omega} \\
&\zeta_k = \frac{\mu_{11}\mu_{22}}{(4\pi)^6 {M^4_k}}
F_1\left(
\frac{m_{h^+_1}^2}{{M^2_k}},
\frac{m_{h^+_2}^2}{{M^2_k}},
0,
0,
\frac{{M_{N_k}}^2}{{M^2_k}},
\frac{m_{k^{\pm\pm}}^2}{{M^2_k}}
\right),
	\label{eq:neutrino_loopfactor}
\end{align}
where $i,j,k$ run over $1$ to $3$,
$m^{\text{diag}}_{\nu}\equiv (m_1,m_2,m_3)$ are the neutrino mass {eigenvalues}, and $U_{{\text{PMNS}}}$ ({Pontecorvo-}Maki-Nakagawa-Sakata matrix~\cite{Maki:1962mu,Pontecorvo:1967fh}) is the mixing matrix to diagonalize the neutrino mass matrix, which is {parametrized} as~\cite{Herrero-Garcia:2014hfa}
\begin{align}
U_{{\text{PMNS}}}=
\left[\begin{array}{ccc} {c_{13}}c_{12} &c_{13}s_{12} & s_{13} e^{-i\delta}\\
 -c_{23}s_{12}-s_{23}s_{13}c_{12}e^{i\delta} & c_{23}c_{12}-s_{23}s_{13}s_{12}e^{i\delta} & s_{23}c_{13}\\
  s_{23}s_{12}-c_{23}s_{13}c_{12}e^{i\delta} & -s_{23}c_{12}-c_{23}s_{13}s_{12}e^{i\delta} & c_{23}c_{13}\\
  \end{array}
\right]
\left[\begin{array}{ccc} 1 & 0 & 0   \\
0 &e^{i\phi/2} & 0\\
 0 & 0 & 1\\
  \end{array}
\right],
\end{align}
with $c_{ij}\equiv \cos\theta_{ij}$ and $s_{ij}\equiv \sin\theta_{ij}$.
$\delta$ and $\phi$ represent the Dirac {$CP$} phase and the Majorana one, respectively.
Here, we treat the two ratios in $F_1$, ${m_{\ell_i}}/{M_k}$ and ${m_{\ell_j}}/{M_k}$ as zero values approximately, which means that the loop factor has no dependence on $i$ or $j$ as $\zeta_k$.

Requirements for the observed neutral profiles depend on the ordering of the neutrino masses, which are normal {[$m_1\,(=0) < m_2 < m_3$]} or inverted {[$m_3\,(=0) < m_1 < m_2$]}.\footnote{More details can be given in Ref.~\cite{Herrero-Garcia:2014hfa} for both of the cases.}
When we consider the normal ordering,
the following relations should hold for realizing the observed neutrino profiles~\cite{Herrero-Garcia:2014hfa},
\begin{align}
(y_{L})_{13} &= (s_{12} c_{23}/(c_{12} c_{13}) + s_{13} s_{23} e^{-i\delta}/c_{13}) (y_{L})_{23}, \notag \\
(y_{L})_{12} &= (s_{12} s_{23}/(c_{12} c_{13}) - s_{13} c_{23} e^{-i\delta}/c_{13}) (y_{L})_{23}, \notag \\
{\left(\frac{(y_{L})_{23}}{2}\right)^2} \omega_{33} &\approx m_3 c^2_{13}s^2_{23}+m_2 e^{i\phi}(c_{12}c_{23}-e^{i\delta}s_{12}s_{13}s_{23})^2, \notag \\
{\left(\frac{(y_{L})_{23}}{2}\right)^2} \omega_{23} &\approx -m_3 c^2_{13}c_{23}s_{23}
+m_2 e^{i\phi}(c_{12}s_{23}+e^{i\delta}c_{23}s_{12}s_{13})(c_{12}c_{23}-e^{i\delta}s_{12}s_{13}s_{23}), \notag \\
{\left(\frac{(y_{L})_{23}}{2}\right)^2} \omega_{22} &\approx m_3 c^2_{13}c^2_{23}+m_2 e^{i\phi}(c_{12}s_{23} {+} e^{i\delta}c_{23}s_{12}s_{13})^2,
	\label{eq:fitting_normal}
\end{align}
where we use $m_e \,{\ll}\, m_\mu, m_\tau$.
In the case of the inverted neutrino mass hierarchy, the conditions are deformed~\cite{Herrero-Garcia:2014hfa},
\begin{align}
(y_{L})_{13} &= - (c_{13} s_{23} e^{-i\delta}/s_{13}) (y_{L})_{23}, \notag \\
(y_{L})_{12} &= + (c_{13} c_{23} e^{-i\delta}/s_{13}) (y_{L})_{23}, \notag \\
{\left(\frac{(y_{L})_{23}}{2}\right)^2} \omega_{33} &\approx m_1 (c_{23}s_{12} + e^{i\delta}c_{12}s_{13}s_{23})^2 + m_2 e^{i\phi}(c_{12}c_{23} - e^{i\delta}s_{12}s_{13}s_{23})^2, \notag \\
{\left(\frac{(y_{L})_{23}}{2}\right)^2} \omega_{23} &\approx m_1 (s_{12}s_{23} - e^{i\delta}c_{12}c_{23}s_{13})(c_{23}s_{12} + e^{i\delta}c_{12}s_{13}s_{23}) \notag \\
&\qquad \qquad \quad + m_2 e^{i\phi}(c_{12}s_{23}+e^{i\delta}c_{23}s_{12}s_{13})(c_{12}c_{23}-e^{i\delta}s_{12}s_{13}s_{23}), \notag \\
{\left(\frac{(y_{L})_{23}}{2}\right)^2} \omega_{22} &\approx m_1 (s_{12}s_{23} - e^{{+i}\delta}c_{12}c_{23}s_{13})^2 + m_2 e^{i\phi}(c_{12}s_{23} {+} e^{i\delta}c_{23}s_{12}s_{13})^2.
	\label{eq:fitting_inverted}
\end{align}
In the numerical analysis in the following {Sec.}~\ref{sec:scanning}, we adopt the latest values of $\theta_{12}$, $\theta_{23}$, $\theta_{13}$, $\Delta m_{21}^2$ and {$\Delta m_{3\ell}^2$} ($\ell = 1$ for {the} normal hierarchy and $\ell = 2$ for {the} inverted one) found in Ref.~\cite{Gonzalez-Garcia:2014bfa}.
The two {$CP$-violating} phases $\delta$ and $\phi$ are treated as free parameters.\footnote{{The best fit values are as follows:
$\sin^2{\theta_{12}} = 0.304$, $\sin^2{\theta_{23}} = 0.452$, $\sin^2{\theta_{13}} = 0.0218$,
$\Delta m_{21}^2 = 7.50 \times 10^{-5}\,\text{eV}^2$, $\Delta m_{31}^2 = +2.457 \times 10^{-3}\,\text{eV}^2$
(in the normal hierarchy);
$\sin^2{\theta_{12}} = 0.304$, $\sin^2{\theta_{23}} = 0.579$, $\sin^2{\theta_{13}} = 0.0219$,
$\Delta m_{21}^2 = 7.50 \times 10^{-5}\,\text{eV}^2$, $\Delta m_{32}^2 = -2.449 \times 10^{-3}\,\text{eV}^2$
(in the inverted hierarchy)~\cite{Gonzalez-Garcia:2014bfa}.
}}

\subsection{Lepton flavor violations and the universality of charged currents}

In the present model, owing to interactions containing charged scalars, new contributions to several {lepton-flavor-violating} processes are found at the tree or the {one loop} level. Also, universalities of charged currents are {threatened} by the vertices.
Some of them are very similar to the case of the Zee-Babu model discussed, {\it e.g.}, in Ref.~\cite{Nebot:2007bc}, while situations are changed in the others.
Here, we briefly summarize important points:

\begin{itemize}
\item $\ell_i \to \ell_j \nu \bar{\nu}$:
like in the Zee-Babu model, the processes receive additional contributions {that} are never found in {the} SM via the exchange of the {singly charged} scalar coupling to the charged leptons $h_1^\pm$, which have an influence on the value of the Fermi constant via {charged-lepton} decay.
If the masses of the {right-hand} neutrinos are small, there is a possibility that the channels $\ell_i \to \ell_j N_1 \bar{N_1}$ via the exchange of $h_2^\pm$ are open and they fake contributions to $\ell_i \to \ell_j \nu \bar{\nu}$.
In this paper, we do not consider such a light $N_1$.
Then, we simply ignore this effect in the following analysis part.
\item $\ell_i^{-} \to \ell_f^{-} \gamma$: like in the Zee-Babu model, these decay sequences (when $m_{\ell_i^{-}} > m_{\ell_f^{-}}$) are radiatively generated at the {one loop} level via the diagrams containing charged scalars which the photon couples to.
In these processes, both of the interactions, $\bar L^c_{L_i} L_{L_j} h^+_1$ and $\bar N_{R_i} e^c_{R_j} h_2^{-}$ {(also {with their Hermitian} conjugations included)} make nonzero contributions to the decay {widths}.
The {doubly charged} scalar $k^{\pm\pm}$ has no contribution {at the {one loop} level} due to the absence of tree-level interactions with the charged leptons.
\item {$\ell_i^{-} \to \ell_j^{+} \ell_k^{-} \ell_l^{-}$}:
being different from the Zee-Babu model, absence of direct interactions between the charged leptons and the {doubly charged} scalar $k^{\pm\pm}$ generates the situation that these processes are induced at the {one loop} level.
Two types of box diagrams contributing to the processes are there: containing (i) $\bar L^c_{L_i} L_{L_j} h^+_1$ interactions; {and containing} (ii) $\bar N_{R_i} e^c_{R_j} h_2^{-}$ ones {(where their {Hermitian} conjugations are also included)}, while two types of photon and $Z$ penguin diagrams are also generated.
Here, the contribution from the penguins is directory related to the dipole diagrams of $\ell_i^{-} \to \ell_f^{-} \gamma$ in the language of branching ratios, {\it e.g.}, when we consider $\mu \to 3e$~\cite{Arganda:2005ji,Toma:2013zsa,Chowdhury:2015sla} {as}
\begin{align}
\text{Br}(\mu \to 3e) \simeq \frac{\alpha_{\text{EM}}}{3\pi} \left( \log\left( \frac{m_\mu^2}{m_e^2} \right) - \frac{11}{4} \right) \text{Br}(\mu \to e\gamma),
\end{align}
where {$\alpha_{\text{EM}}$ is the electromagnetic fine structure constant} and the penguin-type contributions to $\mu \to 3e$ are suppressed compared with those to $\mu \to e\gamma$.
Since the experimental bounds on $\ell_i^{-} \to \ell_f^{-} \gamma$ and $\ell_i^{-} \to \ell_j^{+} \ell_k^{-} \ell_l^{-}$ are roughly comparable, {we} simply ignore the contributions from the penguin-type diagrams in the calculation of the constraints in this category.

\end{itemize}

{Now} we show the explicit forms of constraints on the present model.
We recast the results of the analysis on the Zee-Babu model in Ref.~\cite{Herrero-Garcia:2014hfa} for our case, where
the three $\ell \to \ell \gamma$ processes ($\mu^{-} \to e^{-} \gamma$, $\tau^{-} \to e^{-} \gamma$, $\tau^{-} \to \mu^{-} \gamma$),
the four types of gauge universalities (lepton/hadron, $\mu/e$, $\tau/\mu$, $\tau/e$),
and
the {seven $\ell \to 3\ell$} processes {($\mu^{-} \to e^{+} e^{-} e^{-}$, $\tau^{-} \to e^{+} e^{-} e^{-}$, $\tau^{-} \to e^{+} e^{-} \mu^{-}$, $\tau^{-} \to e^{+} \mu^{-} \mu^{-}$, $\tau^{-} \to \mu^{+} e^{-} e^{-}$, $\tau^{-} \to \mu^{+} e^{-} \mu^{-}$, $\tau^{-} \to \mu^{+} \mu^{-} \mu^{-}$)}
are included.

\begin{table}[t]
\begin{tabular}{c|c|c|c} \hline
Process & $(i,f)$ & Experimental bounds ($90\%$ {C.L.}) & $C_{if}$ \\ \hline
$\mu^{-} \to e^{-} \gamma$ & $(2,1)$ &
	$\text{Br}(\mu \to e\gamma) < 5.7 \times 10^{-13}$ & $1.6 \times 10^{-6}$ \\
$\tau^{-} \to e^{-} \gamma$ & $(3,1)$ &
	$\text{Br}(\tau \to e\gamma) < 3.3 \times 10^{-8}$ & $0.52$ \\
$\tau^{-} \to \mu^{-} \gamma$ & $(3,2)$ &
	$\text{Br}(\tau \to \mu\gamma) < 4.4 \times 10^{-8}$ & $0.7$ \\ \hline
\end{tabular}
\caption{Summary of the coefficient $C_{if}$ in $\ell \to \ell \gamma$ processes and experimental data used in the analysis in~\cite{Herrero-Garcia:2014hfa}.}
\label{tab:Cif}
\end{table}

\begin{table}[t]
\begin{tabular}{c|c} \hline
Type of universality & Experimental bounds ($90\%$ {C.L.}) \\ \hline
{Lepton}/hadron universality & $\sum_{q=d,s,b} |V_{uq}^{\text{exp}}| = 0.9999 \pm 0.0006$ \\
$\mu/e$ universality & $G_{\mu}^{\text{exp}}/G_{e}^{\text{exp}} = 1.0010 \pm 0.0009$ \\
$\tau/\mu$ universality & $G_{\tau}^{\text{exp}}/G_{\mu}^{\text{exp}} = 0.9998 \pm 0.0013$ \\
$\tau/e$ universality & $G_{\tau}^{\text{exp}}/G_{e}^{\text{exp}} = 1.0034 \pm 0.0015$ \\ \hline
\end{tabular}
\caption{Summary of the experimental data on universality of charged currents used in the analysis in~\cite{Herrero-Garcia:2014hfa}.}
\label{tab:gaugeuniv}
\end{table}

\begin{table}[t]
\begin{tabular}{c|c|c|c} \hline
Process & $(i,j,k,l)$ & Experimental bounds ($90\%$ {C.L.}) & $A_{ijkl}$ \\ \hline
$\mu^{-} \to e^{+} e^{-} e^{-}$ & $(2,1,1,1)$ &
	$\text{Br} < 1.0 \times 10^{-12}$ & $2.3 \times 10^{-5}$ \\
$\tau^{-} \to e^{+} e^{-} e^{-}$ & $(3,1,1,1)$ &
	$\text{Br} < 2.7 \times 10^{-8}$ & $0.009$ \\
$\tau^{-} \to e^{+} e^{-} \mu^{-}$ & $(3,1,1,2)$ &
	$\text{Br} < 1.8 \times 10^{-8}$ & $0.005$ \\
$\tau^{-} \to e^{+} \mu^{-} \mu^{-}$ & $(3,1,2,2)$ &
	$\text{Br} < 1.7 \times 10^{-8}$ & $0.007$ \\
$\tau^{-} \to \mu^{+} e^{-} e^{-}$ & $(3,2,1,1)$ &
	$\text{Br} < 1.5 \times 10^{-8}$ & $0.007$ \\
$\tau^{-} \to \mu^{+} e^{-} \mu^{-}$ & $(3,2,1,2)$ &
	$\text{Br} < 2.7 \times 10^{-8}$ & $0.007$ \\
$\tau^{-} \to \mu^{+} \mu^{-} \mu^{-}$ & $(3,2,2,2)$ &
	$\text{Br} < 2.1 \times 10^{-8}$ & $0.008$ \\ \hline
\end{tabular}
\caption{Summary of the coefficient $A_{ijkl}$ in {$\ell \to 3\ell$} processes and experimental data used in the analysis in~\cite{Herrero-Garcia:2014hfa}.}
\label{tab:Aijkl}
\end{table}

\begin{itemize}
\item $\ell_{i}^{-} \to \ell_{f}^{-} \gamma$ processes: in this case, the result of recasting is
\begin{align}
\frac{\left| \sum_{a=1}^3 \left[ (y_L^\dagger)_{af} (y_L)_{ia} \right]^2 \left( I_{1,a} I_{2,a} + I_{1,a}^2 \right) + \sum_{a=1}^3 \left[ (y_R)_{af} (y_R^\dagger)_{ia} \right]^2 \left( I'_{1,a} I'_{2,a} + {I'_{1,a}}^2 \right)  \right|}
{16 m_{h_1^{\pm}}^4 \left| \sum_{a=1}^3 \left( I_{1,a} I_{2,a} + I_{1,a}^2 \right) \right|}
< \frac{C_{if}}{\text{TeV}^4},
	\label{eq:ltolgamma_constraint}
\end{align}
with the loop functions
\begin{align}
I_{1,a} &= {\frac{1}{(4\pi)^2}} {\int_{0}^{1} \!\! dx \int_{0}^{1-x} \!\!\!\!\!\!\! dy}\  \frac{x(2x-1)}{(x+y) m_{h_1^{\pm}}^2 + (1-x-y) m_{\nu_a}^2} \simeq - {\frac{1}{(4\pi)^2}} \frac{1}{36 m_{h_1^{\pm}}^2}, \\
I_{2,a} &= {\frac{1}{(4\pi)^2}} {\int_{0}^{1} \!\! dx \int_{0}^{1-x} \!\!\!\!\!\!\! dy}\  \frac{x(2y-1)}{(x+y) m_{h_1^{\pm}}^2 + (1-x-y) m_{\nu_a}^2} \simeq - {\frac{1}{(4\pi)^2}} \frac{5}{36 m_{h_1^{\pm}}^2}, \\
I'_{1,a} &= {\frac{1}{(4\pi)^2}} {\int_{0}^{1} \!\! dx \int_{0}^{1-x} \!\!\!\!\!\!\! dy}\  \frac{x(2x-1)}{(x+y) m_{h_2^{\pm}}^2 + (1-x-y) M_{N_a}^2}, \\
I'_{2,a} &= {\frac{1}{(4\pi)^2}} {\int_{0}^{1} \!\! dx \int_{0}^{1-x} \!\!\!\!\!\!\! dy}\  \frac{x(2y-1)}{(x+y) m_{h_2^{\pm}}^2 + (1-x-y) M_{N_a}^2},
\end{align}
where we use $m_{\nu_a} \simeq 0$.
{Here, we treat the final-state lepton $\ell^{-}_{f}$ as a massless particle.}
Concrete forms of the integrals are summarized in Appendix~\ref{sec:appendix_loopfunction}.
The dimensionless coefficient $C_{if}$ {representing} the digits in~\cite{Herrero-Garcia:2014hfa} (before recasting) are summarized in {Table}~\ref{tab:Cif}.
The factor $16$ comes from the difference in the coupling convention of the interaction $\bar L^c_{L_i} L_{L_j} h^+_1$ ($\bar L^c_{L_i} L_{L_j} h^+$ in the Zee-Babu model).
These decay processes are {one loop} induced ones in both of the models, and thus the loop factor $1/(4\pi)^2$ in the integrals {is canceled} out in the final form in Eq.~(\ref{eq:ltolgamma_constraint}).
\item Gauge coupling universalities: in this category, recasting is just straightforward by the replacement $h^{\pm} \to h^{\pm}_1$,
\begin{align}
 \left| {\frac{({y}_{L})_{12}}{2}} \right|^2 &<  0.007\ \left(\frac{{m_{h_{1}^\pm}}}{{\rm TeV}}\right)^2
	\quad (\text{lepton/hadron universality}), \notag \\
 \left| \left| {\frac{({y}_{L})_{23}}{2}} \right|^2 - \left| {\frac{({y}_{L})_{13}}{2}} \right|^2 \right| &<  0.024 \ \left(\frac{{m_{h_{1}^\pm}}}{{\rm TeV}}\right)^2
	\quad (\text{$\mu/e$ universality}), \notag \\
 \left| \left| {\frac{({y}_{L})_{13}}{2}} \right|^2 - \left| {\frac{({y}_{L})_{12}}{2}} \right|^2 \right| &<  0.035 \ \left(\frac{{m_{h_{1}^\pm}}}{{\rm TeV}}\right)^2
	\quad (\text{$\tau/\mu$ universality}), \notag \\
 \left| \left| {\frac{({y}_{L})_{23}}{2}} \right|^2 - \left| {\frac{({y}_{L})_{12}}{2}} \right|^2 \right| &<  0.04 \ \left(\frac{{m_{h_{1}^\pm}}}{{\rm TeV}}\right)^2
	\quad \ \,(\text{$\tau/e$ universality}).
	\label{eq:gaugeuniv_constraint}
\end{align}
{The corresponding experimental bounds are summarized in Table~\ref{tab:gaugeuniv}.}
\item {$\ell \to 3\ell$} processes: all of the cases are summarized symbolically as
\begin{align}
\frac{1}{4} {\Big|} \left| A_{\nu} + B_{\nu} \right|^2 + \left| A_{N} + B_{N} \right|^2
-2 \, \text{Re}\left[ A_{N} C_{N}^\ast \right] -2 \, \text{Re}\left[ B_{N} C_{N}^\ast \right]
+ \frac{1}{2} \left| C_N \right |^2  {\Big|^{1/2}}
< \frac{A_{ijkl}}{\text{TeV}^2},
	\label{eq:lto3l_constraint}
\end{align}
with the effective couplings
\begin{align}
A_{\nu} &= (y_L y_L^\dagger)_{ik} (y_L y_L^\dagger)_{jl} \, J_{1,0}, \\
B_{\nu} &= (y_L y_L^\dagger)_{il} (y_L y_L^\dagger)_{jk} \, J_{1,0}, \\
A_{N} &= \sum_{a,b=1}^{3} (y_R)_{al} (y_R)_{bk} (y_R^\dagger)_{ja} (y_R^\dagger)_{ib} \, J_{1,ab}, \\
B_{N} &= \sum_{a,b=1}^{3} (y_R)_{al} (y_R)_{bk} (y_R^\dagger)_{jb} (y_R^\dagger)_{ia} \, J_{1,ab}, \\
C_{N} &= 2 \sum_{a,b=1}^{3} (y_R)_{al} (y_R)_{ak} (y_R^\dagger)_{jb} (y_R^\dagger)_{ib} M_{N_a} M_{N_b} J_{2,ab}.
\end{align}
The loop functions $J_{1,ab}$ and $J_{2,ab}$ are given as
\begin{align}
J_{1,ab} &= \frac{1}{(4\pi)^2} {\int_{0}^{1} \!\! dx \int_{0}^{1-x} \!\!\!\!\!\!\! dy} \  \frac{1-x-y}{x M_{N_{a}}^2 + y M_{N_{b}}^2 + (1-x-y) m_{h_2^{\pm}}^2}, \\
J_{2,ab} &= \frac{1}{(4\pi)^2} {\int_{0}^{1} \!\! dx \int_{0}^{1-x} \!\!\!\!\!\!\! dy} \  \frac{1-x-y}{\left[x M_{N_{a}}^2 + y M_{N_{b}}^2 + (1-x-y) m_{h_2^{\pm}}^2\right]^2},
\end{align}
where the form of $J_{1,0}$ is obtained by the replacements $M_{N_a} \to m_{\nu_a} (\simeq 0)$, $M_{N_b} \to m_{\nu_b} (\simeq 0)$ and $m_{h_2^\pm} \to m_{h_1^\pm}$ as
\begin{align}
J_{1,0} \simeq \frac{1}{2} \frac{1}{(4\pi)^2} \frac{1}{m_{h_1^\pm}^2}.
\end{align}
{Also in this calculation, we treat the final-state leptons $\ell^{+}_{j}$, $\ell^{-}_{k}$ and $\ell^{-}_{l}$ as massless particles.}
Other concrete forms of the integrals are summarized in Appendix~\ref{sec:appendix_loopfunction}.
Here, the dimensionless valuables $A_{ijkl}$ indicate the numerical values (in the analysis in~\cite{Herrero-Garcia:2014hfa} before recasting) and their concrete values are summarized in {Table}~\ref{tab:Aijkl}.
The factor $1/4$ originates from recasting.
The loop suppression factor $1/(4\pi)^2$ should appear in Eq.~(\ref{eq:lto3l_constraint}) since these processes are generated at the {tree level} in the original Zee-Babu model.
\end{itemize}

{Finally, we briefly have a discussion on the LFV via $k^{\pm\pm}$ exchange in our model.
When we focus on the LFV process accompanying two neutrinos, $\ell^{-} \to k^{--} \ell^{+} \to 2 (h_1^{-})^\ast + \ell^{+} \to (2 \ell^{-} + 2 \nu) + \ell^{+}$, where the first step is {one loop} induced and the two $h_1^{-}$ as intermediate states are usually off shell since the mass of {$h_1^{-}$} is bounded from below, where {$m_{h_1^{\pm}}$} should be more than a few TeV; see {Sec.}~\ref{sec:neutrino_fitting} for details.
The order of the loops is the same {as} that of $\ell \to 3\ell$.
But, the {off shellness} suppresses the decay width of this LFV process.
Therefore we can simply ignore this possibility.
}

\section{Valid parameter region with neutrino sector \label{sec:neutrino_fitting}}

\subsection{Form of $y_R$}

Different from the Zee-Babu model, $y_R$ is not a symmetric matrix in general but $y_R^T \, (\zeta M_N) \, y_R \, (\equiv Y_R)$ is {symmetric one}.
Hence, we can utilize fitting relations in the Zee-Babu model with modifications. 
Here, we adopt the following forms in $y_R$ {[and in $y_N$ as shown in Eq.~(\ref{Majonara_mass_form})]} for simplicity,
\begin{align}
y_R=
\left[\begin{array}{ccc} 
\ast & \ast & \ast \\
\ast & a & b \\
\ast & b & c \\
  \end{array}
\right],\quad
M_N={\rm diag}\,(M_{N_1},M_{N_2},M_{N_3}),
	\label{yR_texture}
\end{align}
where $a,b,c$ are arbitrary complex numbers.
The correspondence to the factors $\omega_{ij}$ in Eq.~(\ref{eq:definition_omega}) is as follows:
\begin{align}
\omega_{ij} = m_{\ell_i} (Y_R)_{ij} m_{\ell_j}.
	\label{omega_YR_relation}
\end{align}
Here, because of the mass hierarchy of the charged leptons $m_{\ell_1} \ll m_{\ell_2} < m_{\ell_3}$, the terms in $\omega_{ij}$ including $m_{\ell_1}$ {are} negligible.
Combining this issue with the hierarchy assumed in the {right-hand} neutrinos $M_{N_1} < M_{N_2} < M_{N_3}$, we conclude that the elements of $y_R$ expressed by $\ast$ in Eq.~(\ref{yR_texture}) do not affect the values of $\omega_{ij}$ significantly.
Therefore in the following analysis, we only consider the couplings $a,b,c$ of $y_R$ {in fitting the neutrino profiles}.

Now, we rewrite the relations in Eq.~(\ref{omega_YR_relation}) as follows:
\begin{align}
\frac{\omega_{22}}{m_{\ell_2}^2} &= (Y_{R})_{22} \simeq \left[ a^2 (\zeta_2 M_{N_2}) + b^2 (\zeta_3 M_{N_3}) \right], \notag \\
\frac{\omega_{23}}{m_{\ell_2} m_{\ell_3}} &= (Y_{R})_{23} \simeq \left[ ab \, (\zeta_2 M_{N_2}) + bc \, (\zeta_3 M_{N_3}) \right], \notag \\
\frac{\omega_{33}}{m_{\ell_3}^2} &= (Y_{R})_{33} \simeq \left[ b^2 (\zeta_2 M_{N_2}) + c^2 (\zeta_3 M_{N_3}) \right].
	\label{simultaneous_equation_foryR}
\end{align}
As shown in Eqs.~(\ref{eq:fitting_normal})--(\ref{eq:fitting_inverted}), when we fix the values of {$(y_L)_{23}$}, $\delta$ and $\phi$, the values of $\omega_{22}$, $\omega_{23}$, $\omega_{33}$ {[and also $(y_L)_{12}$, $(y_L)_{13}$]} are automatically determined through the relations.
In each scanning in the following section, we pick up a solution on $a,b,c$ of the above simultaneous equations.

\subsection{Parameter scanning \label{sec:scanning}}

Adopting the structure of $y_R$ in the previous subsection, we execute parameter scans to search for consistent regions in the parameter space.
In this model, $k^{\pm\pm}$ does not contribute to the processes with LFV {significantly}.
Thus, we consider the two possibilities {$m_{k^{\pm\pm}} = 250$ {and} $500\,\text{GeV}$}, while the other two {singly charged} scalars $h_{1^\pm}, h_{2^\pm}$ have a few TeV masses.
Here, we select the four choices: {$m_{h_{1}^\pm} = m_{h_{2}^\pm} = 3,\, 3.5,\, 4,$ {and} $4.8\,\text{TeV}$}.
For brevity, we fix the three {right-hand} neutrinos as follows: $M_{N_1} = m_h/2$, $M_{N_2} = 5\,\text{TeV}$ and $M_{N_3} = 10\,\text{TeV}$, where we mention that $N_1$ is a ``Higgs-portal" dark matter candidate. The mass is assumed to be around the {$125\,\text{GeV}$} Higgs resonant region.
Detailed discussion on this {topic} is found in {Sec.}~\ref{sec:DM calculation}.

We mention that compliance with the relations in the case of the normal hierarchy in Eq.~(\ref{eq:fitting_normal}) or the inverted hierarchy in Eq.~(\ref{eq:fitting_inverted}) leads to the situation that only the {parameters} $\mu_{11}$, $\mu_{22}$, $\delta$, $\phi$, $(y_L)_{23}$ are free to be chosen (in addition to the masses of particles {that} we fixed in the above {paragraph}).
Note that the five matrix elements of $y_R$, namely $(y_R)_{11}$, $(y_R)_{12}$, $(y_R)_{13}$, $(y_R)_{21}$, {and} $(y_R)_{31}$, are ineffective in the determination of the active neutrino profiles, while their {nonzeroness} possibly modifies the strengths of the {lepton-flavor-violating} processes significantly.
{Then} we consider nonzero values in the five couplings for the sake of completeness.
In {each} scanning, we randomly select values of the ten parameters within the corresponding ranges
\begin{align}
\mu_{12} = \mu_{22} \ {(\equiv \mu)} \in [1\,\text{TeV}, 20\,\text{TeV}],\quad
\delta \in [0, 2\pi],\quad \phi \in [0, 2\pi],\quad
(y_L)_{23} \in [-1,1], \notag \\
(y_R)_{ij} \in [-0.01, -0.1] \,\cup\, [0.01, 0.1]\quad
\Big( \text{for } (i,j) = (1,1), (1,2), (1,3), (2,1), (3,1) \Big).
	\label{range_scanning}
\end{align}

Our definition of an allowed point is a set of parameters where all the following requirements are satisfied:
\begin{itemize}
\item
{Observed} values in masses and mixings of the three active neutrinos are generated.
As we mentioned above, realization of this requirement is equal to the compliance with the relations in Eq.~(\ref{eq:fitting_normal}) or (\ref{eq:fitting_inverted}).
\item
{Predictions} for LFVs do not exceed the bounds shown in {Tables}~\ref{tab:Cif} ($\ell \to \ell \gamma$), \ref{tab:gaugeuniv} (gauge universalities), {and} \ref{tab:Aijkl} {($\ell \to 3\ell$)}.
Concretely, evading bounds corresponds to meeting the inequalities in Eqs.~(\ref{eq:ltolgamma_constraint}) {and} (\ref{eq:gaugeuniv_constraint}){--}(\ref{eq:lto3l_constraint}), respectively.
\item
{Fulfilling} the requirements on vacuum stability in Eqs.~{(\ref{eq:vacuumstab_constraint1}){--}(\ref{eq:vacuumstab_constraint3})} and (\ref{form_of_evading_globalminimum}).
In the {last} condition, we adopt the following modified form with physical masses,
\begin{align}
|\mu_{22}|  \lesssim  4 \sqrt{\lambda} \left[m^2_{h_1^\pm} + m^2_{h_2^\pm}+ m^2_{k^{\pm\pm}} \right]^{1/2},
\end{align}
where we simply change the parameters $m_{h_1},\, m_{h_2},\,m_{k}$ to their physical masses and ignore the mass parameter $m_{\Sigma}$.
The value of the right-hand side would be near the original one and it would be useful for roughly estimating this type of {bound}.
In the following analysis, we adopt the initial conditions for {the} quartic couplings,
\begin{align}
\lambda^{(0)}_{h_1} = \lambda^{(0)}_{h_2} = \lambda^{(0)}_{k} = 4\pi,\quad
\lambda = 4\pi.
\end{align}
\item
{Ensuring} perturbativity, all the couplings should be equal to or less than $4\pi$.
\end{itemize}

\begin{table}[tb]
\begin{tabular}{|c|c||c|c||c|} \hline
$m_{k^{\pm\pm}}$ & $(m_{h_1^\pm}, m_{h_2^\pm})$ & {$F_1^{(\zeta_2)}$} & {$F_1^{(\zeta_3)}$} & {Number} of allowed points \\ \hline \hline
     & $(4.8\,\text{TeV}, 4.8\,\text{TeV})$ & $-1.16818$ & $-11.5428$ & {$2804$} \\ \cline{2-5}
$500$ & $(4.0\,\text{TeV}, 4.0\,\text{TeV})$ & $-2.14198$ & $-20.0287$ & {$422$} \\ \cline{2-5}
$\text{GeV}$ & $(3.5\,\text{TeV}, 3.5\,\text{TeV})$ & $-3.30195$ & $-29.6525$ & {$89$} \\ \cline{2-5}
     & $(3.0\,\text{TeV}, 3.0\,\text{TeV})$ & $-5.37518$ & $-46.0932$ & {$16$} \\ \hline
     & $(4.8\,\text{TeV}, 4.8\,\text{TeV})$ & $-1.21466$ & $-11.9695$ & {$1644$} \\ \cline{2-5}
$250$ & $(4.0\,\text{TeV}, 4.0\,\text{TeV})$ & $-2.24746$ & $-20.9444$ & {$190$} \\ \cline{2-5}
$\text{GeV}$ & $(3.5\,\text{TeV}, 3.5\,\text{TeV})$ & $-3.49141$ & $-31.2321$ & {$31$} \\ \cline{2-5}
     & $(3.0\,\text{TeV}, 3.0\,\text{TeV})$ & $-5.74153$ & $-49.0141$ & {$2$} \\ \hline 
\end{tabular}
\caption{The numbers of consistent points in the normal hierarchy obtained by parameter scanning.
{$F_1^{(\zeta_2)}$ and $F_1^{(\zeta_3)}$ are the loop functions in $\zeta_2$ and $\zeta_3$ shown in Eq.~(\ref{eq:neutrino_loopfactor}), respectively.}
Kindly refer to the main body of this subsection for details of this scanning.}
\label{tab:parameter_scan}
\end{table}

\begin{figure}[t]
\begin{center}
\includegraphics[width=0.40\columnwidth]{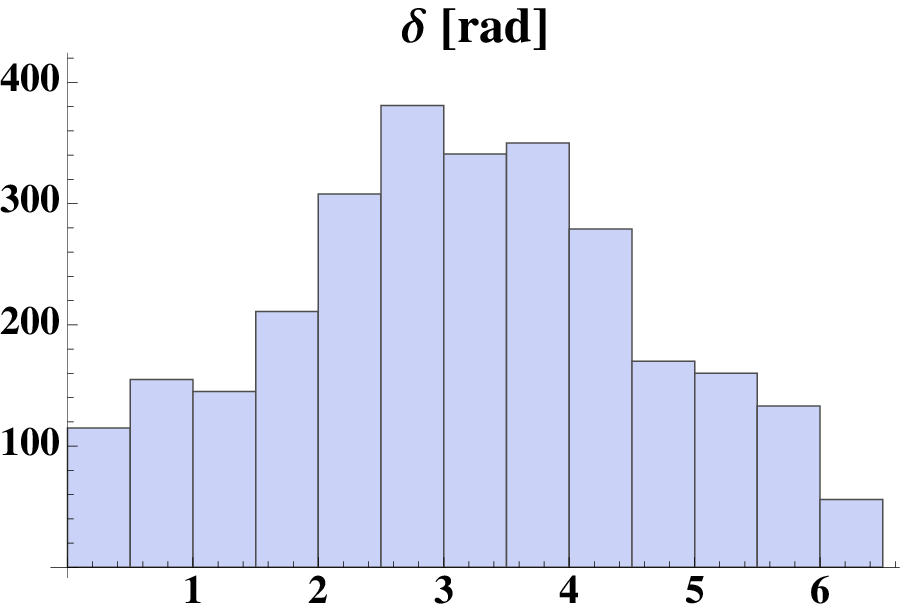}
\includegraphics[width=0.40\columnwidth]{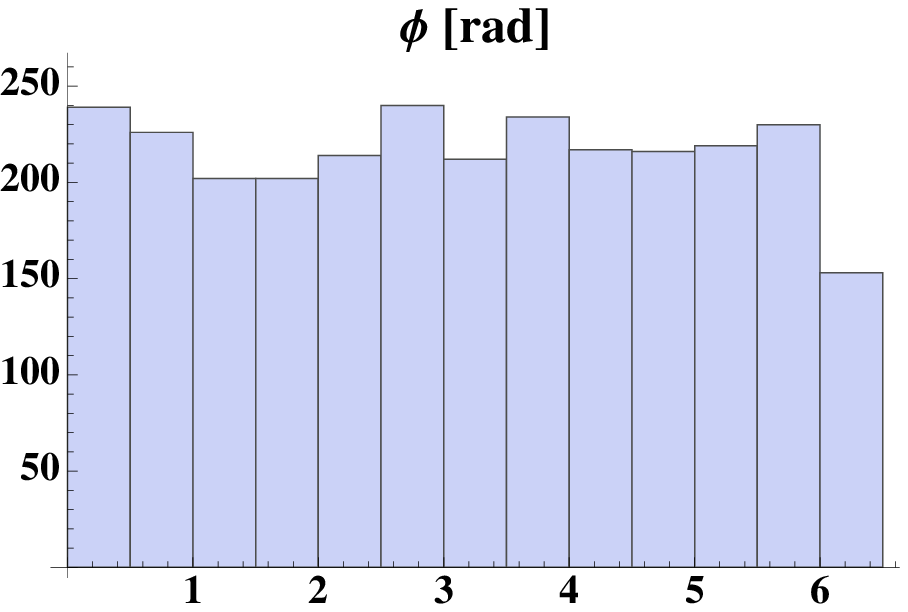}
\\[10pt]
\includegraphics[width=0.40\columnwidth]{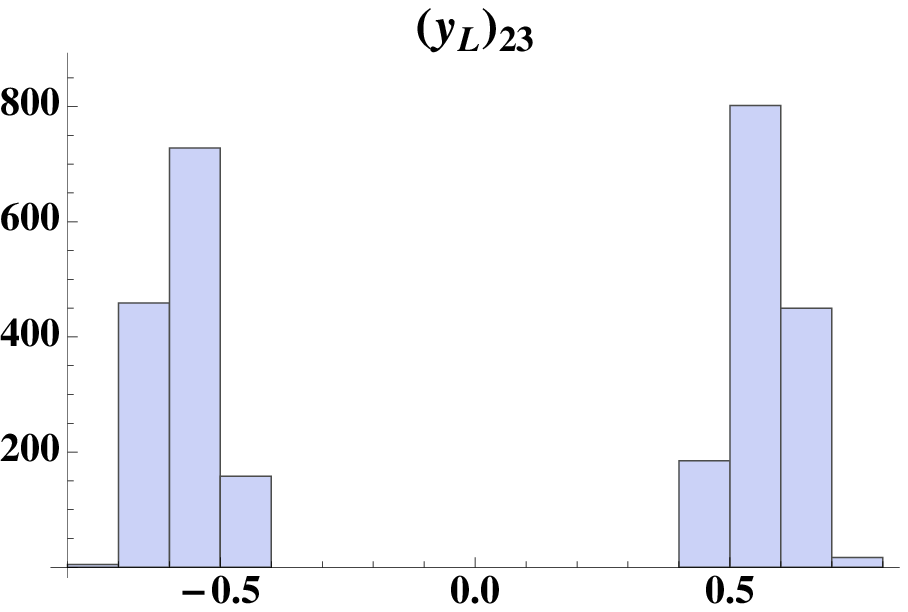}
\includegraphics[width=0.40\columnwidth]{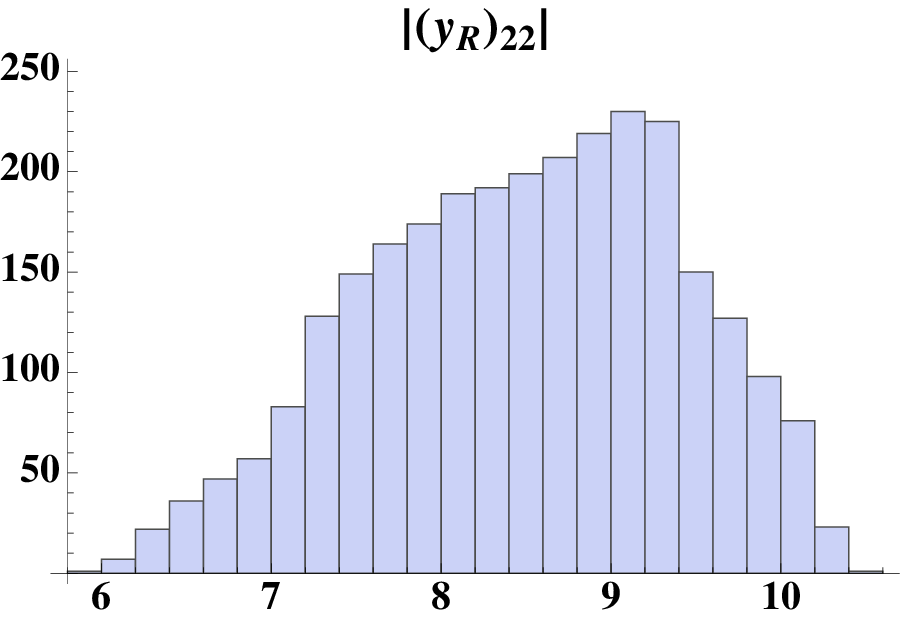}
\\[10pt]
\includegraphics[width=0.40\columnwidth]{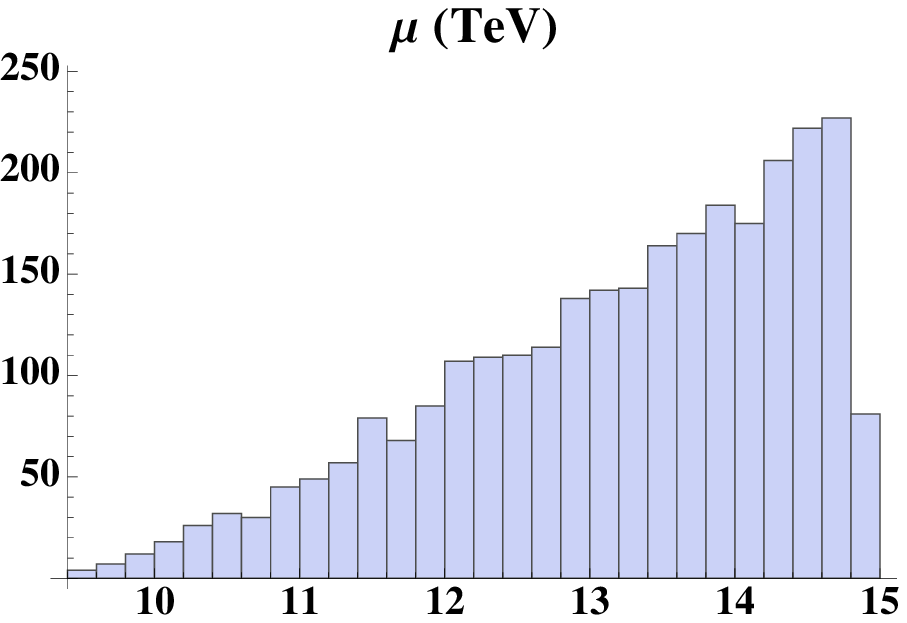}
   \caption{Histograms showing distributions of $\delta$, $\phi$, $(y_L)_{23}$, $|(y_R)_{22}|$ and $\mu$ of the result of a scanning with the charged scalar masses $m_{k^{\pm\pm}} = 500\,\text{GeV},\, m_{h_1^\pm} = m_{h_2^\pm} = 4.8\,\text{TeV}$.
   The total number of the consistent data points is {$2804$}.}
   \label{fig:scandata}
\end{center}
\end{figure}

\begin{figure}[t]
\begin{center}
\includegraphics[width=0.40\columnwidth]{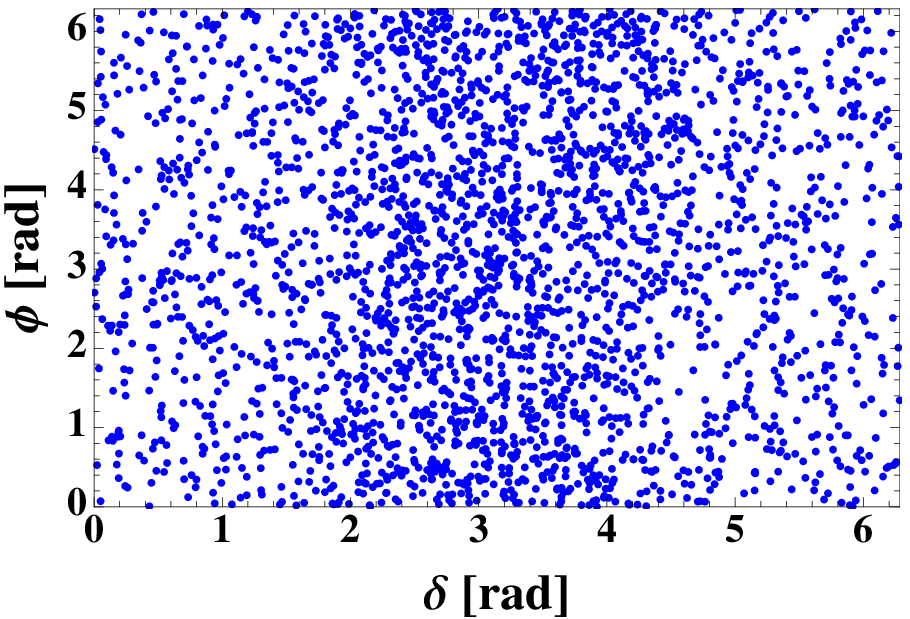}
\\[10pt]
\includegraphics[width=0.42\columnwidth]{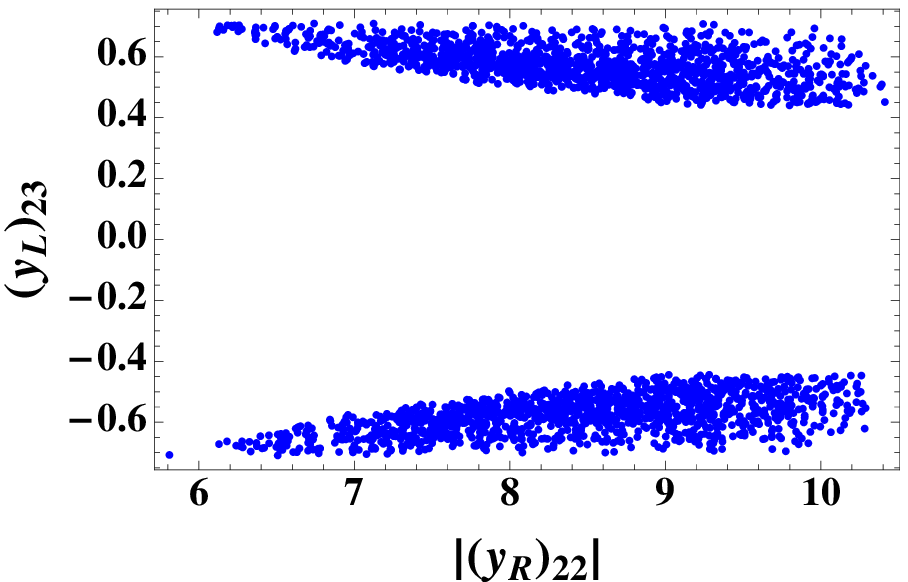} \
\includegraphics[width=0.40\columnwidth]{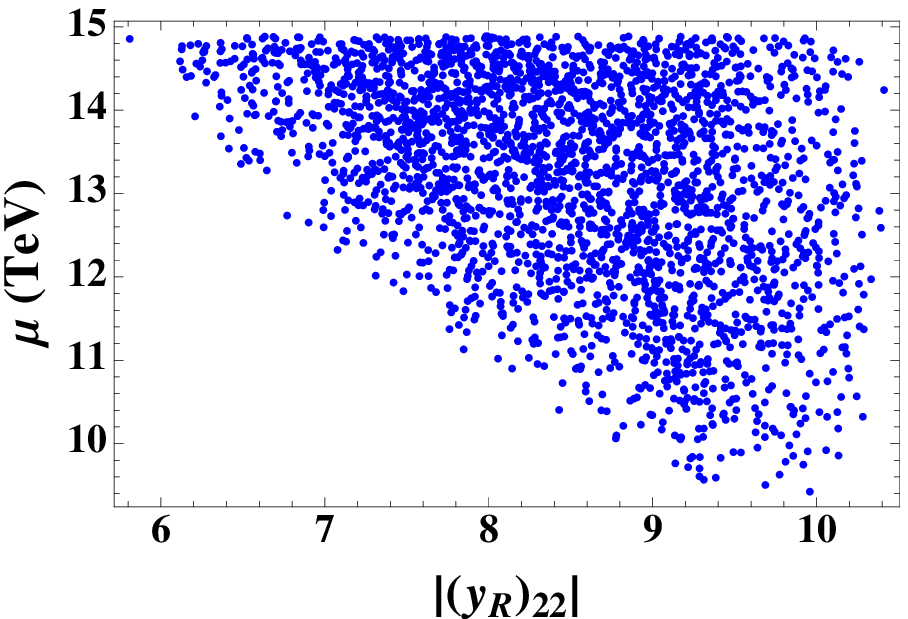}
   \caption{
Correlations between $\phi$ and $\delta$, $(y_L)_{23}$ and $|(y_R)_{22}|$, {and} $\mu$ and $|(y_R)_{22}|$ are shown.
The {data set} is the same {as} in Fig.~\ref{fig:scandata}.
}
   \label{fig:scandata_cor}
\end{center}
\end{figure}

Our result in the normal hierarchy is summarized in {Table}~\ref{tab:parameter_scan}.
In each of the eight combinations of the charged scalar masses, we randomly scan $10^5$ points in the ten parameters, where the range we consider is shown in Eq.~(\ref{range_scanning}).
Apart from the previous work~\cite{Hatanaka:2014tba}, the result says that the mass of the {doubly charged} scalar can be light around a few hundred GeV, whereas the other two {singly charged} ones should be heavy (at least) around a few TeV.
In this scenario, it is very hard to produce $h_1^\pm$ and $h_2^\pm$ in colliders.
On the other hand, detecting $k^{\pm\pm}$ could be a reasonable option for probing this model in present and future collider experiments.\footnote{{Lower bounds at $95\%$ CL on $m_{k^{\pm\pm}}$ via $8\,\text{TeV}$ data were provided by the ATLAS group in Ref.~\cite{ATLAS:2014kca} as $374\,\text{GeV}$, $402\,\text{GeV}$, $438\,\text{GeV}$ when assuming a $100\%$ branching ratio to $e^{\pm} e^{\pm}$, $e^{\pm} \mu^{\pm}$, $\mu^{\pm} \mu^{\pm}$ pairs, respectively.
In our model, the decay sequence $k^{\pm\pm} \to \ell^{\pm} \ell^{\pm}$ is one-loop induced and many parameters contribute.
In this paper, we skip to have detailed discussions on prospects in colliders.
}}

Next, we look into properties of the parameters in the allowed region.
Here, we only show the case of $m_{k^{\pm\pm}} = 500\,\text{GeV},\, m_{h_1^\pm} = m_{h_2^\pm} = 4.8\,\text{TeV}$ as an example, where we checked {that} other cases are not {so different}.
In Fig.~\ref{fig:scandata}, we display the distributions of $\delta$, $\phi$, $(y_L)_{23}$, $|(y_R)_{22}|$ and $\mu$ based on {$2804$} data points of the allowed region as histograms.\footnote{{The value of $(y_R)_{22}$ is complex in general obtained as a part of a solution of the simultaneous equations in Eq.~(\ref{simultaneous_equation_foryR}).}}
{The correlations between $\phi$ and $\delta$, $(y_L)_{23}$ and $|(y_R)_{22}|$, {and} $\mu$ and $|(y_R)_{22}|$ are also shown in Fig.~\ref{fig:scandata_cor}.}
In the present {three loop} scenario, the loop factor {$1/(4\pi)^{6}$} in the neutrino masses tends to suppress the realized masses {very} much.
Thus, at least a part of the parameters related with the masses would be sufficiently large.
In the {following}, we investigate details.

\begin{itemize}
\item
We can see that this model could give some preference to $\delta$ around $\pi$.
On the other hand, $\delta$ not around $\pi$ is also realized.
Few {trends are} seen in the distribution of $\phi$.
\item
A typical digit of the absolute value of $(y_L)_{23}$ is {in $0.5 \sim 0.6$}, which would be greater compared with other radiative neutrino models in {one} or two loop level.
Since $|\omega_{22}| \simeq |\omega_{23}| \simeq |\omega_{33}|$, the other two nonzero components of $y_L$, $(y_L)_{12}$ and $(y_L)_{13}$, {tend to} have almost the same order of magnitude.
\item
Like the Zee-Babu model, via the relation $|\omega_{22}| \simeq |\omega_{23}| \simeq |\omega_{33}|$, the three components of the effective symmetric Yukawa couplings $Y_R$ meet the definite hierarchy,
\begin{align}
{|(Y_R)_{33}| : |(Y_R)_{23}| : |(Y_R)_{22}|} \sim
\frac{m_{\mu}^2}{m_{\tau}^2} : \frac{m_{\mu}}{m_{\tau}} : 1,
\end{align}
which means that the original {$|(y_R)_{22}|$} tends to hold a significantly large value.
The peak of the distribution is around {$9$}, which is still rather small compared with the
perturbative upper bound $4\pi$.
Because of this characteristic, the masses of the {singly charged} scalars $m_{h_1^\pm}$ and $m_{h_2^\pm}$ should be greater than around $3\,\text{TeV}$ to circumvent the bounds.
\item
The {common} coefficients of the trilinear terms {$\mu$} should be large {at} around {$14 \sim 15\,\text{TeV}$} to compensate {for} the suppression factor in the neutrino masses.
From the perturbativity in the couplings {$\lambda_{11}$ in the form of the trilinear coupling $\mu_{11}$ in Eq.~(\ref{form_mu11}) and $y_N$ found in the masses of the { right-hand} neutrinos in Eq.~(\ref{Majonara_mass_form})}, $v'$ cannot be so small.
A reasonable value of $v'$ is $\mathcal{O}(1)\,\text{TeV}$.
\end{itemize}

Finally, we briefly comment on the possibility with the inverted mass hierarchy.
In our model, like the Zee-Babu model, the components of the active neutrino mass matrix $(m_\nu)_{ab}$ {contain} the charged lepton masses{, and also} the Majorana masses with the assumed mass hierarchy $M_{N_{1}} < M_{N_{2}} < M_{N_{3}}$.
{Then} the normal hierarchy would be preferable.
Within our search among $10^5$ points in the inverted case, even in the choice of $(m_{k^{\pm\pm}}, m_{h_1^\pm}, m_{h_{2}^\pm}) = (500\,\text{GeV}, 4.8\,\text{TeV}, 4.8\,\text{TeV})$, {no} solution is found.

\section{Constraint from LHC Higgs search \label{sec:LHCglobalfit}}

\begin{table}[tb]
\begin{tabular}{|c||c|c|c|} \hline
Process & ATLAS & CMS & Reference \\ \hline
$h \to \gamma\gamma$ & $1.17^{+0.27}_{-0.27}$ & $1.12 \pm 0.24$ & \cite{Aad:2014eha,Khachatryan:2014jba} \\
$h \to ZZ^\ast \to 4\ell$ & $1.44^{+0.40}_{-0.33}$ & $1.00 \pm 0.29$ & \cite{Aad:2014eva,Khachatryan:2014jba} \\
$h \to WW^\ast \to \ell\nu\ell\nu$ & $1.09^{+0.23}_{-0.21}$ & $0.83 \pm 0.21$ & \cite{ATLAS:2014aga,Khachatryan:2014jba} \\
$h \to b \bar{b}$ & $0.5^{+0.4}_{-0.4}$ & $0.84 \pm 0.44$ & \cite{Aad:2014xzb,Khachatryan:2014jba} \\
$h \to \tau \bar{\tau}$ & $1.4^{+0.4}_{-0.4}$ & $0.91 \pm 0.28$ & \cite{ATLAS_Higgs_tautau,Khachatryan:2014jba} \\ \hline
\end{tabular}
\caption{Summary of the latest LHC Higgs search data as (combined) signal strengths in the five Higgs decay patterns ($h \to \gamma\gamma, h \to ZZ^\ast \to 4\ell, h \to WW^\ast \to \ell\nu\ell\nu, h \to b \bar{b}, h \to \tau \bar{\tau}$). The analyses are based on the {data sets} accumulated in the LHC first run, whose details of the ATLAS and the CMS are $4.5$ -- $4.7 \, \text{fb}^{-1}\,(7\,\text{TeV}) + 20.3 \, \text{fb}^{-1}\,(8\,\text{TeV})$, $5.1 \, \text{fb}^{-1} \, (7 \, \text{TeV}) + 19.7 \, \text{fb}^{-1} \, (8 \, \text{TeV})$, respectively.}
\label{tab:LHC_Higgs_latestdata}
\end{table}

In this part, we evaluate constraints on the parameter space of scalars in this model by use of the latest results of LHC Higgs searches by the ATLAS and CMS experiment groups.
First, we describe the method we use for global analysis.
Like in the papers~\cite{Giardino:2012dp,Carmi:2012in,Ellis:2012hz}, we adopt the following form of signal strength of the single Higgs production channel with the subsequent Higgs decay to the particles $f$,
\begin{align}
\mu_f = \frac{\sigma_{\text{total}}}{\sigma_{\text{total}}^{\text{SM}}} \times \frac{\text{Br}_{h \to f}}{\text{Br}_{h \to f}^{\text{SM}}},
\label{def_signalstrength}
\end{align}
where $\sigma_{\text{total}}$ ($\sigma_{\text{total}}^{\text{SM}}$) represents the total production cross section of a Higgs boson in this model (SM), respectively.
The Higgs branching ratios to the particles $f$ of {the} SM and this model are discriminated by {having or not having the superscript SM}.

Here, note that in our scenario, the observed Higgs boson is a mixture of the $SU(2)_L$ doublet $\Phi$ and the singlet $\Sigma_0$ as shown in Eq.~(\ref{eq:mass_weak}) and no additional colored particle is introduced, which means the absence of new contributions to the Higgs production via the gluon fusion process.
Since the SM gauge bosons and quarks do not couple to $\Sigma_0$, the ratio of the total cross sections is easily calculated as
\begin{align}
\frac{\sigma_{\text{total}}}{\sigma_{\text{total}}^{\text{SM}}}
=
\cos^2{\alpha}.
\end{align}

Whereas, evaluating the ratio of the branding ratios is rather complicated.
First, we look at the following decomposition,
\begin{align}
\frac{\text{Br}_{h \to f}}{\text{Br}_{h \to f}^{\text{SM}}}
=
\left( \frac{\Gamma_{h}}{\Gamma^{\text{SM}}_h} \right)^{-1} \times
\frac{\Gamma_{h \to f}}{\Gamma^{\text{SM}}_{h \to f}},
\end{align}
where $\Gamma_{h}^{(\text{SM})}$ represents the corresponding Higgs total width.
In the Higgs decay, the presence of the charged scalars in our model ($h_1^\pm, h_2^\pm, k^{\pm\pm}$) also modifies the partial decay widths of the photon-associated decay processes, $h \to \gamma\gamma$ and $h \to Z \gamma$, in addition to the doublet-singlet mixing effect.
As we discussed in {Sec.}~{\ref{sec:scanning}}, the two {singly charged} scalars tend to be very massive {at} around a few TeV at least.
In such a situation, we can completely neglect the effect via the two particles and then only the contribution from $k^{\pm\pm}$ is included as a loop effect in the following part.

The ratio of the total widths is rewritten as follows:
\begin{align}
\frac{\Gamma_{h}}{\Gamma^{\text{SM}}_h}
=
\text{Br}^{\text{SM}}_{h \to \text{SM\,others}} \times \frac{\Gamma_{h \to \text{SM\,others}}}{\Gamma_{h \to \text{SM\,others}}^{\text{SM}}} +
\text{Br}^{\text{SM}}_{h \to \gamma\gamma} \times \frac{\Gamma_{h \to \gamma\gamma}}{\Gamma^{\text{SM}}_{h \to \gamma\gamma}} +
\text{Br}^{\text{SM}}_{h \to Z\gamma} \times \frac{\Gamma_{h \to Z\gamma}}{\Gamma^{\text{SM}}_{h \to Z\gamma}} +
\frac{\Gamma_{h \to \text{inv}}}{\Gamma^{\text{SM}}_{h}},
\end{align}
where $\Gamma_{h \to \text{inv}}$ expresses the Higgs partial decay width to invisible pairs, which is written as
\begin{align}
\Gamma_{h \to \text{inv}} = \Gamma_{h \to GG} + \Gamma_{h \to N_{R_1} \bar{N}_{R_1}}.
\end{align}
Concrete forms of partial decay width are summarized in Appendix~\ref{sec:appendix_width}.

$\text{Br}^{\text{SM}}_{h \to \text{SM\,others}}$ can be described as $1 - \text{Br}^{\text{SM}}_{h \to \gamma\gamma} - \text{Br}^{\text{SM}}_{h \to Z\gamma}$ very precisely.
In the following analysis, we use the values $\text{Br}^{\text{SM}}_{h \to \gamma\gamma} = 2.28 \times 10^{-3}, \, \text{Br}^{\text{SM}}_{h \to Z\gamma} = 1.54 \times 10^{-3}$ and $\Gamma^{\text{SM}}_{h} = 4.07\,\text{MeV}$ in $m_h = 125\,\text{GeV}$ reported by the LHC Higgs Cross Section Working Group~\cite{Heinemeyer:2013tqa}.
The ratios of the partial decay widths are described with the help of the formulas for $h \to \gamma\gamma$ and $h \to Z\gamma$ in Refs.~\cite{Ellis:1975ap,Shifman:1979eb,Djouadi:2005gi,Carena:2012xa,Chen:2013vi} as
\begin{align}
\frac{\Gamma_{h \to f}}{\Gamma^{\text{SM}}_{h \to f}} &= \cos^2{\alpha} \quad (f \in (\text{others in SM})) \quad \to \quad \frac{\Gamma_{h \to \text{SM\,others}}}{\Gamma_{h \to \text{SM\,others}}^{\text{SM}}} = \cos^2{\alpha}, \\
\frac{\Gamma_{h \to \gamma\gamma}}{\Gamma^{\text{SM}}_{h \to \gamma\gamma}}
&=
\left| \cos{\alpha} + \frac{1}{2} \frac{v^2}{m_{k^{\pm\pm}}^2} Q_k^2 C_{hkk} \frac{A_0^{\gamma\gamma}(\tau_k)}{A_1^{\gamma\gamma}(\tau_W) + N_C Q_t^2 A_{1/2}^{\gamma\gamma}(\tau_t)} \right|^2, \\
\frac{\Gamma_{h \to Z\gamma}}{\Gamma^{\text{SM}}_{h \to Z\gamma}}
&=
\left| \cos{\alpha} - \frac{v^2}{m_{k^{\pm\pm}}^2} \left( 2 Q_k g_{Zkk} \right) C_{hkk} \frac{A_0^{Z\gamma}(\tau_k,\lambda_k)}{v A_{\text{SM}}^{Z\gamma}} \right|^2,
\end{align}
with effective couplings
\begin{align}
C_{hkk} = \cos{\alpha} \lambda_{\Phi k} - \sin{\alpha} \left( \frac{v'}{v} \right) \lambda_{\Sigma k},\quad
g_{Zkk} = - Q_k \left( \frac{s_W}{c_W} \right)
	\label{effective_couplings_oneloopHiggs}
\end{align}
and the form factor
\begin{align}
A_{\text{SM}}^{Z\gamma} = \frac{2}{v} \left[ \cot{\theta_W} A_{1}^{Z\gamma}(\tau_W,\lambda_W) + N_C
\frac{(2Q_t)(T_3^{(t)} - 2Q_t s_W^2)}{s_W c_W} A_{{1/2}}^{Z\gamma}(\tau_t,\lambda_t) \right].
\end{align}
$N_C\,(=3)$, $Q_t\,(=2/3)$, $Q_k\,(=2)$, $T_3^{(t)}\,(=1/2)$, $c_W$ and $s_W$ are the QCD color factor for quarks, the electric charges of the top quark, {the doubly charged scalar in unit} of the positron's one, the weak isospin of the top quark, {and} the cosine and the sine of the Weinberg angle $\theta_W$, respectively.
The loop factors take the following forms,
\begin{align}
A_1^{\gamma\gamma}(x)
&=
-x^2 \left[ 2x^{-2} + 3x^{-1} + 3(2x^{-1}-1) f(x^{-1}) \right], \\
A_{1/2}^{\gamma\gamma}(x)
&=
2x^2 \left[ x^{-1} + (x^{-1} -1) f(x^{-1}) \right], \\
A_{0}^{\gamma\gamma}(x)
&=
-x^2 \left[ x^{-1} - f(x^{-1}) \right], \\
A_{1}^{Z\gamma}(x,y)
&=
4 (3 - \tan^2{\theta_W}) I_2(x,y) + \left[ (1+2x^{-1}) \tan^2{\theta_W} - (5+2x^{-1}) \right] I_1(x,y), \\
A_{1/2}^{Z\gamma}(x,y)
&=
I_1(x,y) - I_2(x,y), \\
A_{0}^{Z\gamma}(x,y)
&=
I_1(x,y),
\end{align}
with the functions
\begin{align}
I_1(x,y) &= \frac{xy}{2(x-y)} + \frac{x^2y^2}{2(x-y)^2} \left[ f(x^{-1}) - f(y^{-1}) \right] + \frac{x^2y}{(x-y)^2} \left[ g(x^{-1}) - g(y^{-1}) \right], \\
I_2(x,y) &= - \frac{xy}{2(x-y)} \left[ f(x^{-1}) - f(y^{-1}) \right].
\end{align}
In the above formulas, forms of the input variables to the loop factors $\tau_i$ and $\lambda_i$ are defined as fractions by the Higgs boson mass ($m_h$) or the Z boson mass ($m_Z$)
\begin{align}
\tau_i = \frac{4 m_i^2}{m_h^2}, \quad \lambda_i = \frac{4 m_i^2}{m_Z^2} \quad (i=t,W,k).
\end{align}
The two ratios usually take values above one ($m_{h} \leq 2 m_i,\, m_{Z} \leq 2 m_i$).
The two functions $f(z)$ and $g(z)$ ($z \equiv x^{-1}\ \text{or}\ y^{-1}$) are formulated as
\begin{align}
f(z) &=
\begin{cases}
\arcsin^2{\sqrt{z}} & \text{for } z \leq 1, \\
-\frac{1}{4} \left[ \log\left( \frac{1+\sqrt{1-z^{-1}}}{1-\sqrt{1-z^{-1}}} \right) -i\pi \right]^2 & \text{for } z > 1,
\end{cases} \\
g(z) &=
\begin{cases}
\sqrt{z^{-1} - 1} \arcsin{\sqrt{z}} & \text{for } z \leq 1, \\
\frac{\sqrt{1-z^{-1}}}{2} \left[ \log\left( \frac{1+\sqrt{1-z^{-1}}}{1-\sqrt{1-z^{-1}}} \right) -i\pi \right] & \text{for } z > 1,
\end{cases}
\end{align}
where the situation $m_{h} \leq 2 m_i,\, m_{Z} \leq 2 m_i$ corresponds to $z \leq 1$.

\begin{figure}[t]
\begin{center}
\includegraphics[width=0.40\columnwidth]{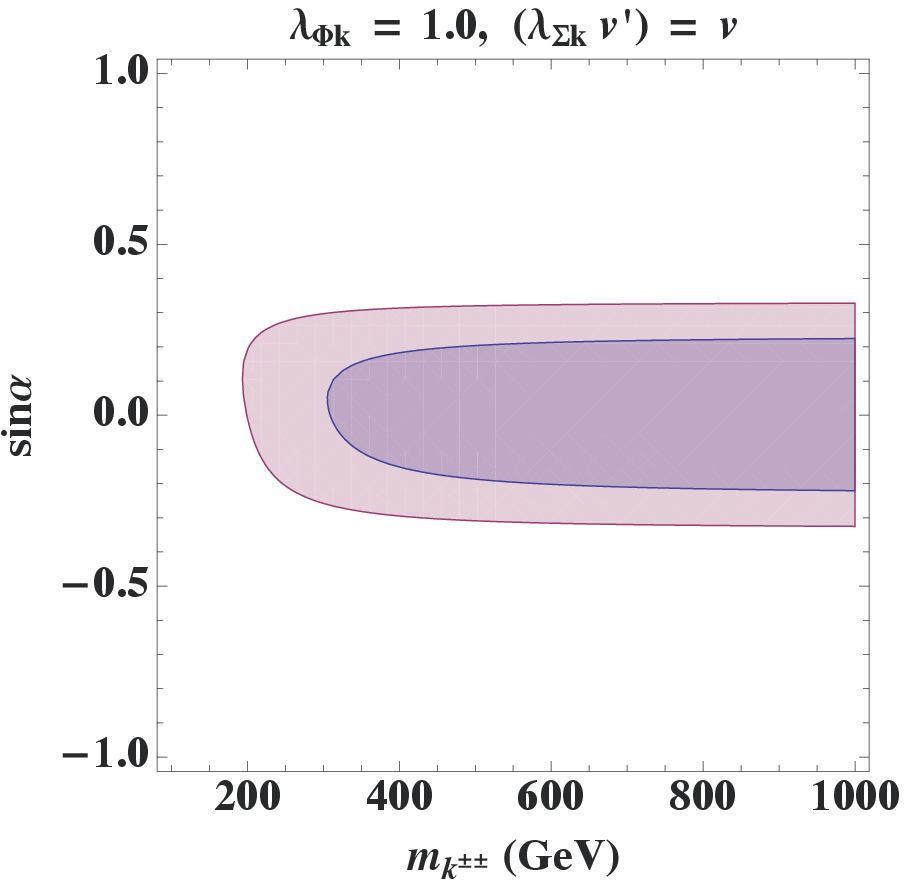}
\includegraphics[width=0.40\columnwidth]{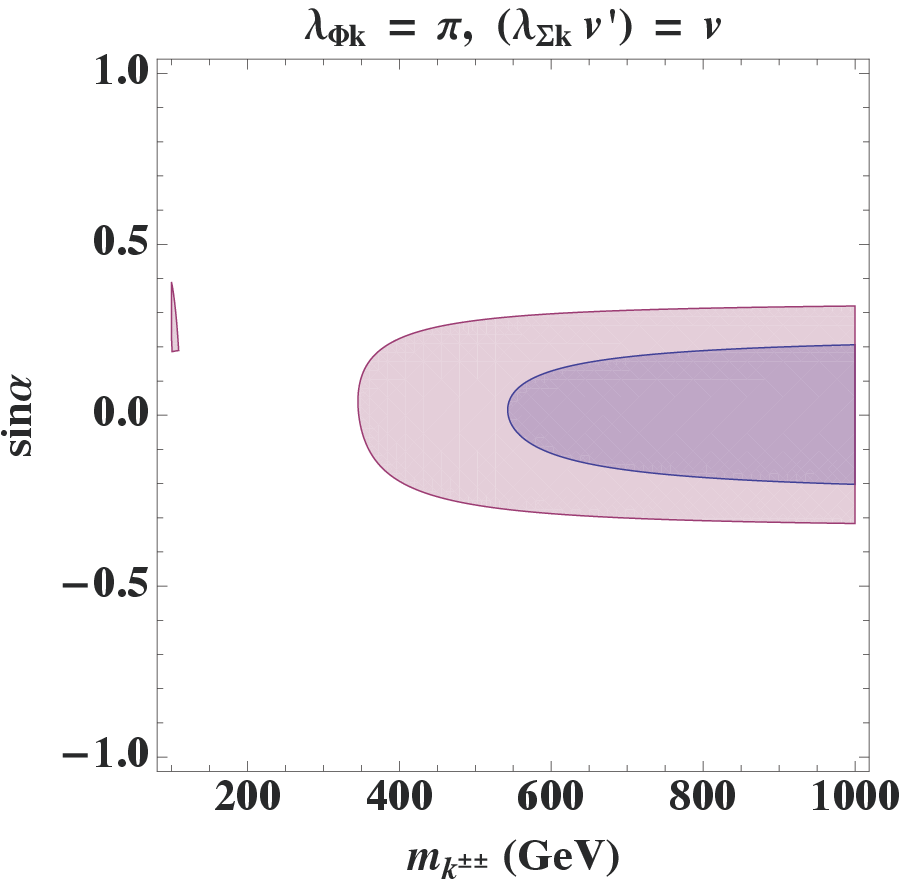}
\includegraphics[width=0.40\columnwidth]{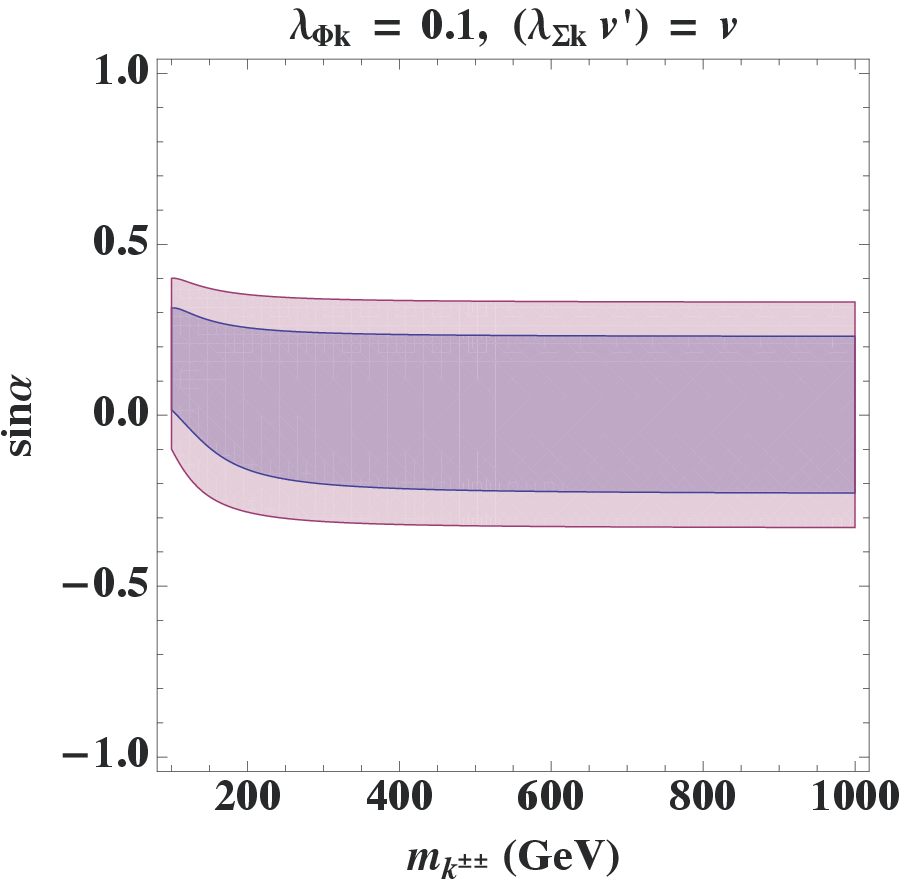}
   \caption{The $1\sigma$ (blue) and $2\sigma$ (red) allowed regions of the global analysis based on the LHC data summarized in {Table}~\ref{tab:LHC_Higgs_latestdata}. From top left to bottom, we choose the parameters $(\lambda_{\Phi k}, \lambda_{\Sigma k} v') = (1.0, v),\, (\pi, v),\, (0.1, v)$, respectively.}
   \label{fig:LHCHiggscontour_1}
\end{center}
\end{figure}

\begin{figure}[t]
\begin{center}
\includegraphics[width=0.40\columnwidth]{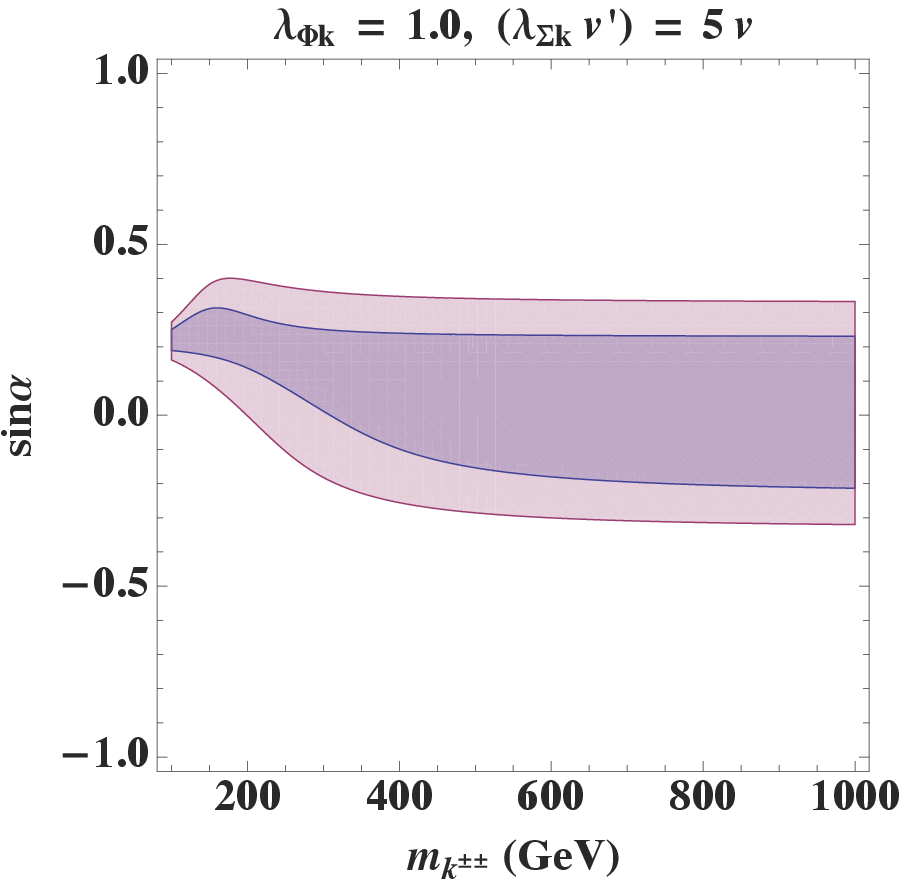}
\includegraphics[width=0.40\columnwidth]{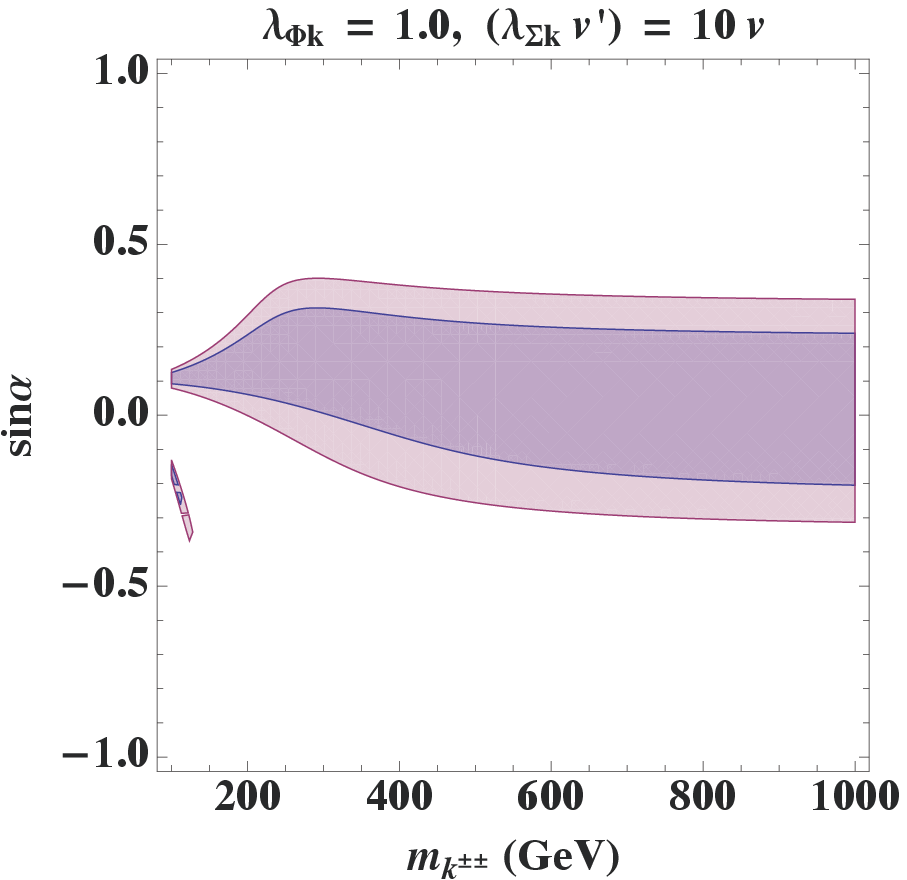}
   \caption{The $1\sigma$ (blue) and $2\sigma$ (red) allowed regions of the global analysis based on the LHC data summarized in {Table}~\ref{tab:LHC_Higgs_latestdata}. From left to right, we choose the parameters $(\lambda_{\Phi k}, \lambda_{\Sigma k} v') = (1.0, 5 v),\, (1.0, 10 v)$, respectively.}
   \label{fig:LHCHiggscontour_2}
\end{center}
\end{figure}

{To estimating the} consistent parameter region with the latest results of the Higgs search at {the} LHC, we define the $\chi^2$ valuable by use of the signal strength in Eq.~(\ref{def_signalstrength}) as follows:
\begin{align}
\chi^2 =
\sum_{f=\gamma\gamma,ZZ^\ast,WW^\ast,\atop b\bar{b},\tau\bar{\tau}\,(\text{ATLAS})}
\left( \frac{\mu_{f} - \hat{\mu}_f}{\hat{\sigma}_f} \right)^2 +
\sum_{f=\gamma\gamma,ZZ^\ast,WW^\ast,\atop b\bar{b},\tau\bar{\tau}\,(\text{CMS})}
\left( \frac{\mu_{f} - \hat{\mu}_f}{\hat{\sigma}_f} \right)^2.
\end{align}
Here, we take the results of the five Higgs decay channels reported by the ATLAS and the CMS experiments into consideration, which are $h \to \gamma\gamma, h \to ZZ^\ast \to 4\ell, h \to WW^\ast \to \ell\nu\ell\nu, h \to b \bar{b}, h \to \tau \bar{\tau}$ summarized in {Table}~\ref{tab:LHC_Higgs_latestdata}.
The two hatted symbols $\hat{\mu}_f$ and $\hat{\sigma}_f$ represent the corresponding central value and error, respectively.
We assumed that each of ten experimental inputs follows the Gaussian distribution and there are no correlations among them.
Also, when an error is asymmetric, we adopt its simple average as an input value of the corresponding $\hat{\sigma}_f$ for analysis.
These simplifications {are} justified for our purpose of roughly estimating survived regions of the parameter space of this model.

In the following analysis, for simplicity, we focus on a reasonable situation {in which} the mass of the DM is around $m_h/2$ where the Higgs invisible channel to a pair of the DMs is near the threshold and negligible.
Detailed discussions on the DM candidate {are given in Sec.}~\ref{sec:DM calculation}.
Now, apparently, we see the {five parameters} $m_{k^{\pm\pm}}$, $\lambda_{\Phi k}$, $\lambda_{\Sigma k}$, $v'$ and $\alpha$ govern the signal strengths of the Higgs boson.
But in fact, as shown in Eq.~(\ref{effective_couplings_oneloopHiggs}), {$\lambda_{\Sigma k}$} and $v'$ contribute to the Higgs physics only as the combination of $\lambda_{\Sigma k} v'$.
Thus, four independent degrees of freedom are relevant in total.

We fix the two parameters $\lambda_{\Phi k}$ and $\lambda_{\Sigma k} v'$ in each of the following global fits for simplicity.
Here, we consider the five possibilities, $(\lambda_{\Phi k}, \lambda_{\Sigma k} v') = (1.0, v)$, $(\pi, v)$,  $(0.1, v)$, $(1.0, 5v)$ and $(1.0, 10v)$ and search for their global minima on the remaining two variables $(m_{k^{\pm\pm}}, \sin{\alpha})$ at first.
The positions of {each} corresponding minimum are $(m_{k^{\pm\pm}}, \sin{\alpha}) = ({74.7}\,\text{GeV}, -0.0759)$, $(109\,\text{GeV}, -0.0774)$, {$(65.0\,\text{GeV}, 0.102)$}, $(80.3\,\text{GeV}, -0.0768)$ and $(87.5\,\text{GeV}, -0.0804)$ with $\chi^2_{\text{min}} \simeq 5.08$ (commonly in all the cases), respectively.
The $1\sigma$ and $2\sigma$ boundaries are defined as the positions with $\chi^2 = \chi^2_{\text{min}} + \Delta \chi^2_{1\sigma, 2\sigma}$, where the values of $\Delta \chi^2_{1\sigma, 2\sigma}$ are calculated by the cumulative distribution function of the $\chi^2$-distribution with two degrees of freedom as $\Delta\chi^2_{1\sigma} = 2.296$, $\Delta \chi^2_{2\sigma} = 6.180$.
The results are shown in Figs.~\ref{fig:LHCHiggscontour_1}{--}\ref{fig:LHCHiggscontour_2}.
{Small disconnected regions are found possibly due to accidental cancellations when $(\lambda_{\Phi k}, \lambda_{\Sigma k} v') = (\pi, v)$ in Fig.~\ref{fig:LHCHiggscontour_1} and $(\lambda_{\Phi k}, \lambda_{\Sigma k} v') = (1.0, 10v)$ in Fig.~\ref{fig:LHCHiggscontour_2}.}
When we consider the {doubly charged} scalars with $m_{k^{\pm\pm}} = 500\,\text{GeV}$, the range of $\alpha$
\begin{align}
|\sin{\alpha}| \ {\lesssim} \  {0.3}
\end{align}
is roughly allowed within $2\sigma$ confidence levels, where actual upper bounds depend on contents of the two prefixed parameters.

\section{Issues on $N_{R_1}$ dark matter \label{sec:DM calculation}}

In this section, we have discussions on dark matter related issues of this model.
After the breakdown of the new global $U(1)$ symmetry, an accidental $Z_2$ symmetry still remains, which ensures the existence of a dark matter candidate.
Under the assignment of the $U(1)$ charge in {Table}~\ref{tab:1}, $N_{R_i}\,(i=1,2,3)$ and $h^\pm_2$ hold negative parities, where the lightest one among them gets to be absolutely stable.
After considering the aptness discussed in {Sec.}~\ref{sec:scanning} that $h^{\pm}_2$ (and also $h^\pm_1$) should be sufficiently heavy {at} around a few TeV to evade the constraints, {\it e.g.}, via the processes with LFV, the lightest {right-hand} neutrino, namely $N_{R_1}${,} in our setup is usually stabilized by the symmetry and plays a significant role as {dark matter}.

An important point of our dark sector is in the mechanism for generating Majorana mass terms of $N_R$ {that} originates from the spontaneous global $U(1)$ breaking accompanying a nonzero VEV of the field $\Sigma_0$.
Through the mixing between $\Sigma_0$ and the Higgs doublet $\Phi$, $N_{R_1}$ can communicate with the SM particles, where the two neutral scalars $h$ and $H$ work as mediators.
Here, we expect that {a suitable amount} of dark matter relics are left in the {Universe} through resonant effects via the $h$ or $H$ pole, where the mass of $N_{R_1}$ is fixed around $m_h/2$ or $m_H/2$, respectively.

The relic abundance and the spin-independent cross section for direct detection of the candidate $N_{R_1}$ are calculable by following a standard method.
{In the following part, we use the {shorthand} notation $\chi$ for showing the DM $N_{R_1}$ (in the mass eigenstate as a Majorana fermion).}
Within the {freeze-out} approximation, the present-day relic density is evaluated as~\cite{Griest:1990kh}
\begin{align}
\Omega_\chi h^2 \simeq \frac{1.07 \times 10^9\,\text{GeV}^{-1}}{\sqrt{g_\ast(x_f)} M_{\text{Pl}} J(x_f)},
\end{align}
where $\Omega_\chi$, $h\,(\simeq 0.7)$, $M_{\text{Pl}}$ and $g_\ast$ express the present-day energy density of $\chi$, the present-day scaled Hubble parameter, the Planck mass and the number of the relativistic degrees of freedom, respectively.
Note that the latest value of $\Omega_{\chi} h^2$ reported by the Planck experiment is $0.1196 \pm 0.0031\,(68\%\,\text{{C.L.}})$~\cite{Ade:2013zuv}.
The valuable $x_f$ is defined by the dark matter mass $m_\chi$ and the temperature at the {freeze-out} $T_f$ as $m_\chi/T_f$.
The efficiency of the annihilation after the {freeze-out} is described through the integral $J$:
\begin{align}
J(x_f) = \int_{x_f}^{\infty} dx \frac{\langle \sigma v_{\text{rel}} \rangle}{x^2},
\end{align}
where $\langle \sigma v_{\text{rel}} \rangle$ stands for the {thermally averaged} annihilation cross section multiplied with the relativistic relative velocity $v_{\text{rel}}$.
In this work, we adopt the sophisticated way for taking thermal average relativistically discussed in Refs.~\cite{Gondolo:1990dk,Edsjo:1997bg}, where $\langle \sigma v_{\text{rel}} \rangle$ is estimated as
\begin{align}
\langle \sigma v_{\text{rel}} \rangle
=
\frac{\displaystyle \int_{4m_\chi^2}^{\infty} ds \frac{\sqrt{s-4m_\chi^2}}{16} W_{\chi\chi} K_1\left( \frac{\sqrt{s}}{T} \right)}{\displaystyle m_\chi^4 T \left[ K_2 \left( \frac{m_\chi}{T} \right)\right]^2}
=
\frac{\displaystyle \int_{4m_\chi^2}^{\infty} ds {\sqrt{s-4m_\chi^2}} W_{\chi\chi} K_1\left( \frac{\sqrt{s}}{m_\chi} x \right)}{\displaystyle 16 m_\chi^5 x^{-1} \left[ K_2 \left( x \right)\right]^2},
\end{align}
where $x$ is defined as $m_\chi/T$ (like $x_f$) at the temperature $T$, $K_{1,2}$ are the modified Bessel functions of the second kind of order $1$ and $2$, respectively, and $W_{\chi\chi}$ is a Lorentz invariant variable describing the cross section multiplied with the Lorentz invariant flux factor $4 E_\chi^2 v_{\text{rel}}$.
$W_{\chi\chi}$ is formulated in the center of mass system with the integration over the solid angle as
\begin{align}
W_{\chi\chi} = \sum_{f} \frac{1}{32\pi^2} \sqrt{1- \frac{4m_f^2}{s}} \int d\Omega \left| \overline{\mathcal{M}}(\chi {\bar{\chi}} \to f \bar{f}) \right|^2
\end{align}
where we sum over {all} possible two-body final states consisting of the same particle $f$.
The detail of the amplitudes {is} found in Appendix~\ref{sec:appendix_amplitude}.
In the following numerical calculation, we simply use the fixed reasonable values $x_f = 20$ and $g_\ast(x_f) = 100$ throughout {the} calculations for brevity.

Our evaluation of {the} $\chi$-nucleon spin-independent cross section is based on the discussions in {Refs.}~\cite{Hisano:2010ct,Abe:2015rja}.
We consider the following effective Lagrangian for calculating the cross section at the leading order,
\begin{align}
\mathcal{L}_{\text{eff}}
=
\sum_{{q}=u,d,s} f_q m_q \bar{\chi} \chi \bar{q} q - \frac{\alpha_s}{4\pi}
f_G \bar{\chi} \chi G^{a}_{\mu\nu} G^{a\mu\nu},
\end{align}
where $q$, $m_q$, $\alpha_s${,} and $G^{a}_{\mu\nu}$ represent the corresponding quark fields, the quark masses, the QCD coupling strength{,} and the field strength of the gluon, respectively.
The coefficients $f_q$ ($f_G$) determine the effective interactions between the quarks (gluon) and {the DM $\chi$}.
The corresponding values in our model are
\begin{align}
f_q = f_G = \frac{1}{2} \left( \frac{m_\chi}{vv'} \right) \left( -\frac{1}{m_h^2} + \frac{1}{m_H^2} \right) \cos{\alpha} \sin{\alpha}.
\end{align}

The spin-independent cross section with the target nucleus $T$ is formulated by use of the spin-independent coupling of $\chi$ with nucleon $f_N\,(N=p,n)$ as
\begin{align}
\sigma_{\text{SI}}^{T} = \frac{4}{\pi} \mu_{T\chi}^2 \left| n_p f_p + n_n f_n \right|^2,
\end{align}
where $\mu_{T\chi}$ is the reduced mass among the nucleus and the dark matter {is} defined as $\mu_{T\chi} \equiv m_T m_\chi/(m_T + m_\chi)$. $m_T$ shows the mass of the target nucleus $T$, which contains $n_p$ and $n_n$ numbers of protons ($p$) and neutrons ($n$), respectively.
The effective $\chi$-nucleon ($N$) coupling $f_N$ can be written down by the coefficients {of the effective operators in $\mathcal{L}_{\text{eff}}$} ($f_q, f_G$) and matrix elements ($f_{Nq}, f_{NG}$) as 
\begin{align}
f_N/m_N = \sum_{q=u,d,s} f_q f_{Nq} + \frac{2}{9} f_G f_{NG},
\end{align}
with the concrete forms of the matrix elements,
\begin{align}
\bra{N} m_q \bar{q} q \ket{N} = f_{Nq} m_N,\quad
1 - \sum_{q=u,d,s} f_{Nq} = f_{NG}{,}
\end{align}
where the value of $f_{NG}$ is calculable through the latter relation by {the} use of $f_{Nq}$.
Like in Ref.~\cite{Abe:2015rja}, we adopt the following default values in the program {\tt micrOMEGAs}~\cite{Belanger:2013oya}: $f_{nu} = 0.0110, \, f_{nd} = 0.0273, \, f_{ns} = 0.0447$.

The latest bound on the spin-independent scattering process was reported by the LUX experiment as an upper limit on the spin-independent (elastic) DM-nucleon cross section, which is approximately $10^{-45}\,\text{cm}^2$ (when $m_\chi \sim 10^2\,\text{GeV}$) with the $90\%$ confidence level~\cite{Akerib:2013tjd}.
Here, we choose the {neutron} as the nucleon for putting {a} bound on our parameter space via the LUX result. 
Now, the (spin-independent) $\chi$-{neutron} cross section is calculated with ease as
\begin{align}
\sigma_{\text{SI}}^{N=n} &= \frac{4}{\pi} \mu_{n\chi}^2 \left| m_n \left( \sum_{q=u,d,s} f_{nq} + \frac{2}{9} f_{nG} \right) f_q \right|^2 \notag \\
&=
\frac{1}{\pi} \mu_{n\chi}^2 m_n^2 \, C^2 \left( \frac{m_\chi \cos{\alpha} \sin{\alpha}}{vv'} \right)^2
\left( -\frac{1}{m_h^2} + \frac{1}{m_H^2} \right)^2,
\end{align}
with
\begin{align}
C = \frac{2}{9} + \frac{7}{9} \sum_{q=u,d,s} f_{nq} \simeq 0.287,
\end{align}
where similar calculations were done, {\it e.g.}, in Refs.~\cite{Baek:2011aa,Baek:2012se}.

\begin{figure}[t]
\centering
\includegraphics[width=0.40\columnwidth]{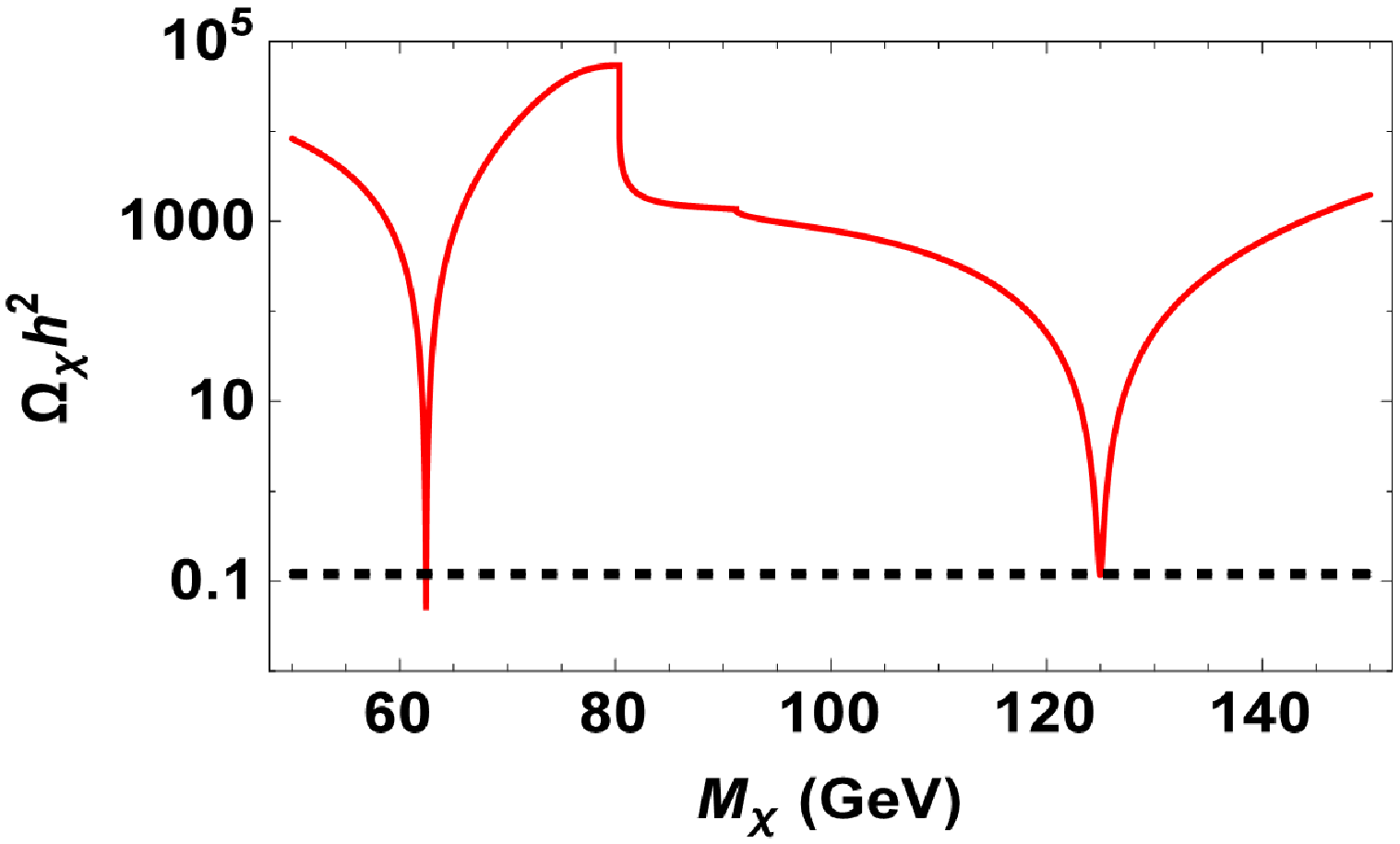}
\quad
\includegraphics[width=0.40\columnwidth]{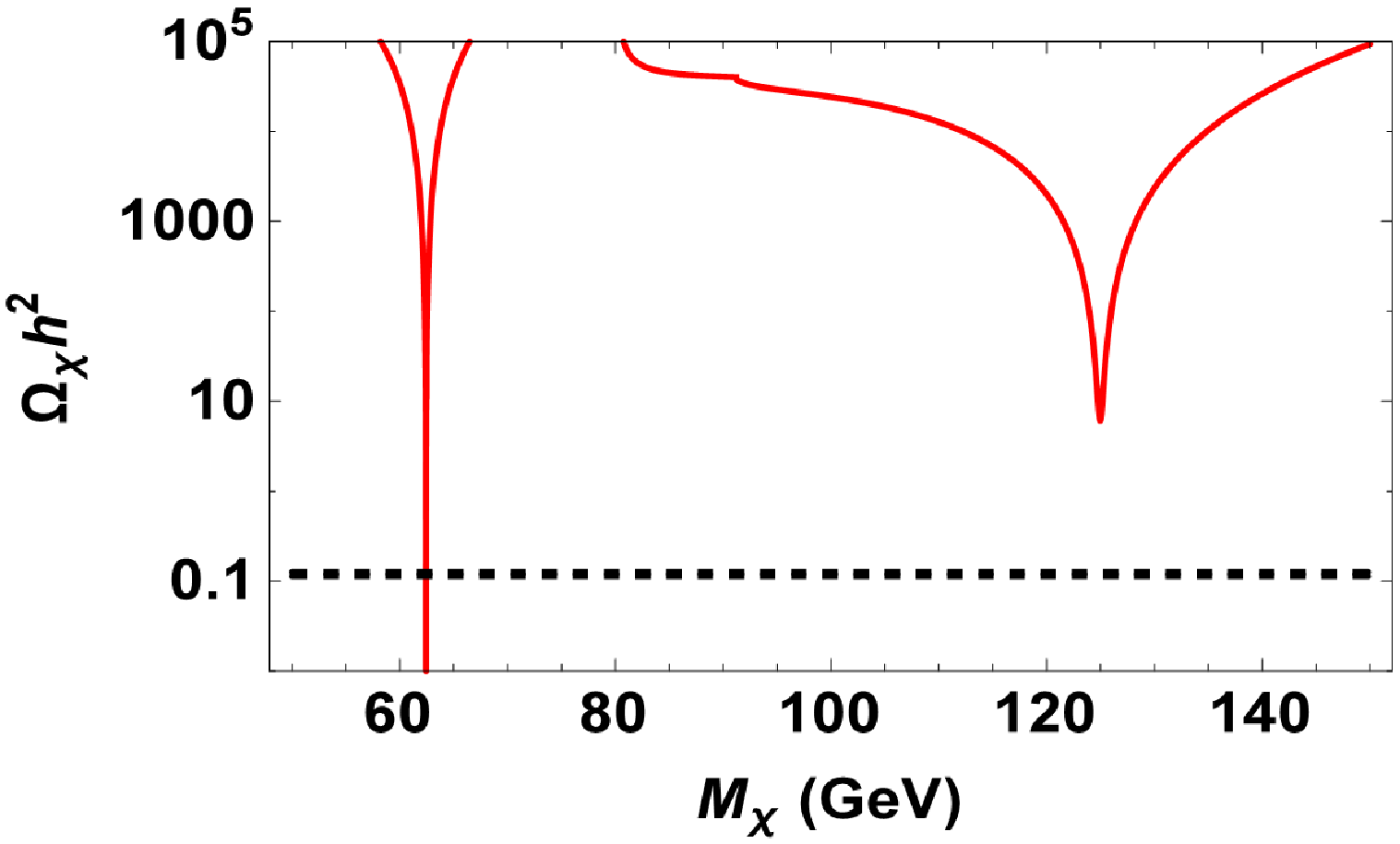}
   \caption{Realized present-day relic densities of the dark matter candidate $\chi$ are shown as functions of $m_\chi$. The other effective parameters are fixed as $\sin{\alpha} = 0.4,\, v' = 800\,\text{GeV},\, m_H = 250\,\text{GeV},\, m_{k^{\pm\pm}} = 500\,\text{GeV}$ (left panel) and $\sin{\alpha} = 0.5,\, v' = 5\,\text{TeV},\, m_H = 250\,\text{GeV},\, m_{k^{\pm\pm}} = 500\,\text{GeV}$ (right panel) for demonstration {purposes}, respectively.
The area inside the two horizontal dashed lines (where the splitting is almost invisible) suggests the $2\sigma$ consistent region with the Planck data {[$0.1196 \pm 0.0031\,(68\%\,\text{C.L.})$~\cite{Ade:2013zuv}]}.
In the current setup, the relic density sharply drops around the two resonant regions around $m_h/2 \simeq 62.5\,\text{GeV}$ and $m_H/2 \simeq 125\,\text{GeV}$, where in the latter case, shown in the right panel, dropping is not sufficient for obtaining a proper amount of dark matter even around $m_H/2$ at present.
We neglect the three- and four-body final states via virtual $W$ and $Z$ boson decays, which gives sizable modifications near the thresholds for producing gauge boson pairs~\cite{Cline:2012hg,Cline:2013gha}, since our interest is only around $m_h/2$ and $m_H/2$ where this correction is expected to be insignificant.
}
   \label{fig:reliccurve}
\end{figure}

\begin{figure}[t]
\centering
\includegraphics[width=0.495\columnwidth]{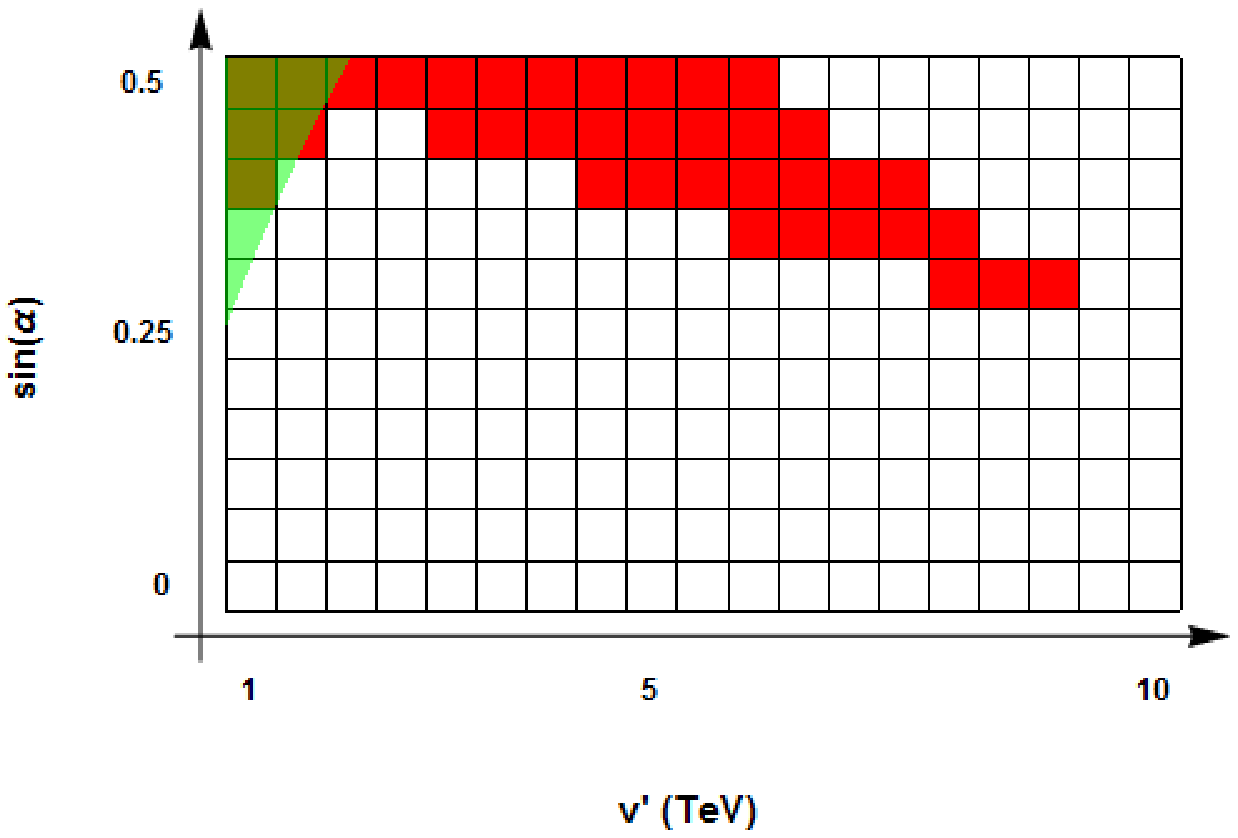}
\includegraphics[width=0.495\columnwidth]{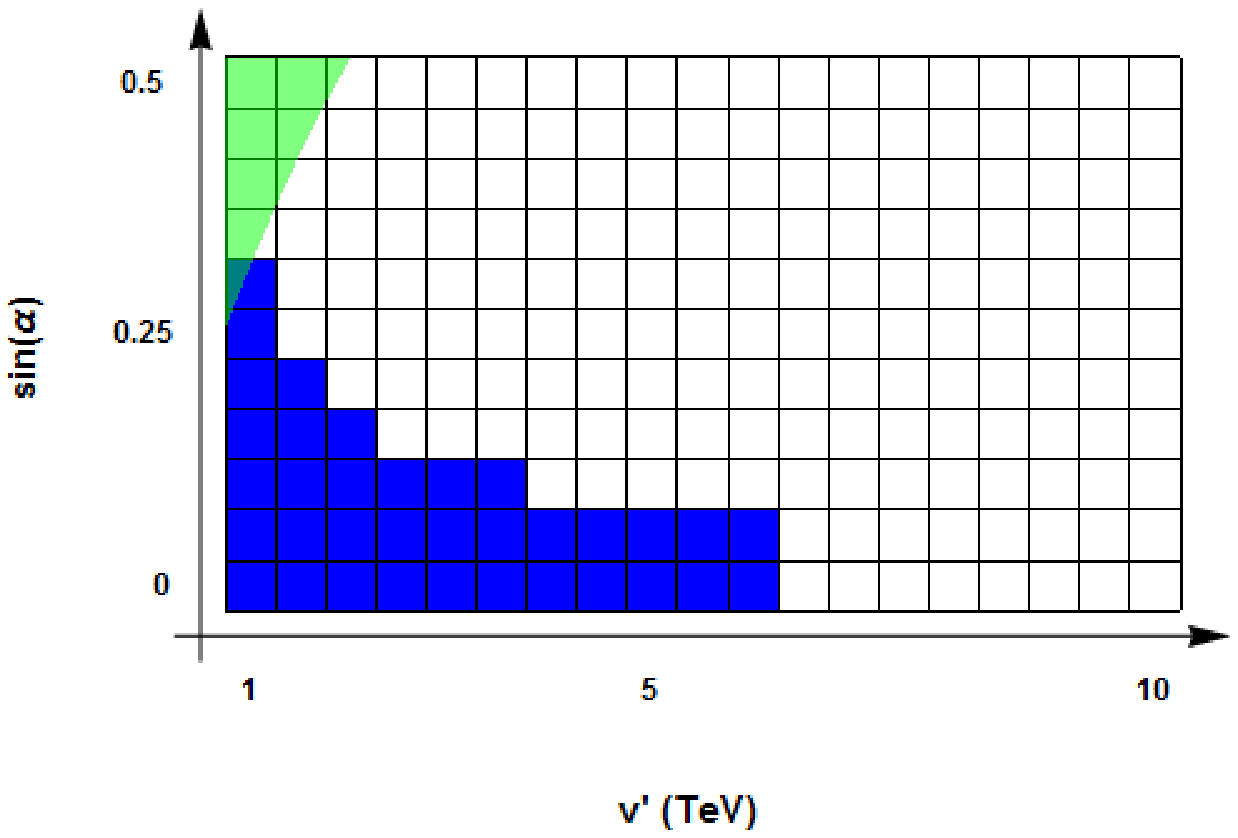}
   \caption{The matrix plots {indicate} suitable choices of the two parameters $v'$ and $\sin{\alpha}$ when we consider the situation of $m_H = 250\,\text{GeV},\, m_{k^{\pm\pm}} = 250\,\text{GeV}$.
We obtain the observed relic density in the red (blue) region in the left (right) panel when $m_\chi$ is around $m_h/2\,\simeq 62.5\,\text{GeV}$ ($m_H/2\,\simeq 125\,\text{GeV}$), respectively.
The green region is excluded via excess of the invisible decay channel of the observed Higgs boson.
{No excluded region is found by the direct detection in the shown parameter range.}
}
   \label{fig:DM_matrixplot_MH250GeV_Mk250GeV}
\end{figure}

\begin{figure}[t]
\centering
\includegraphics[width=0.495\columnwidth]{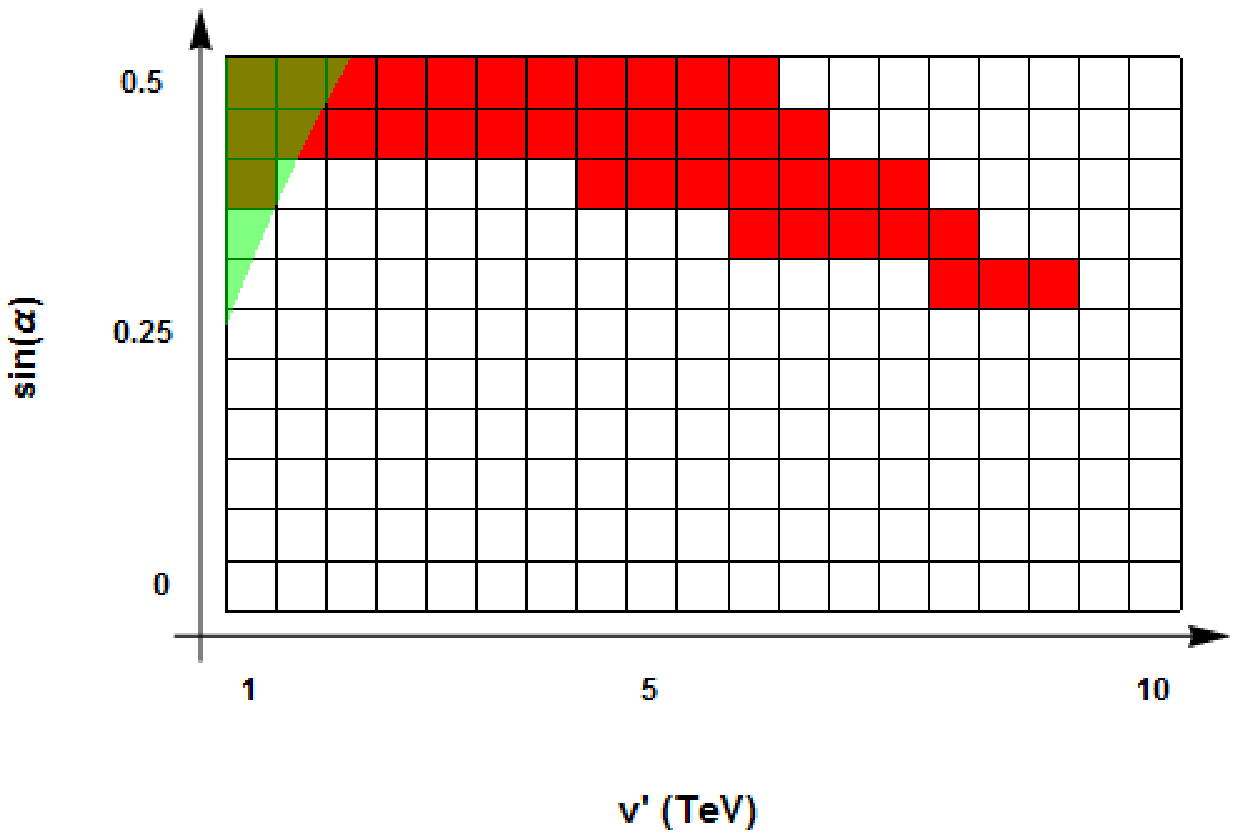}
\includegraphics[width=0.495\columnwidth]{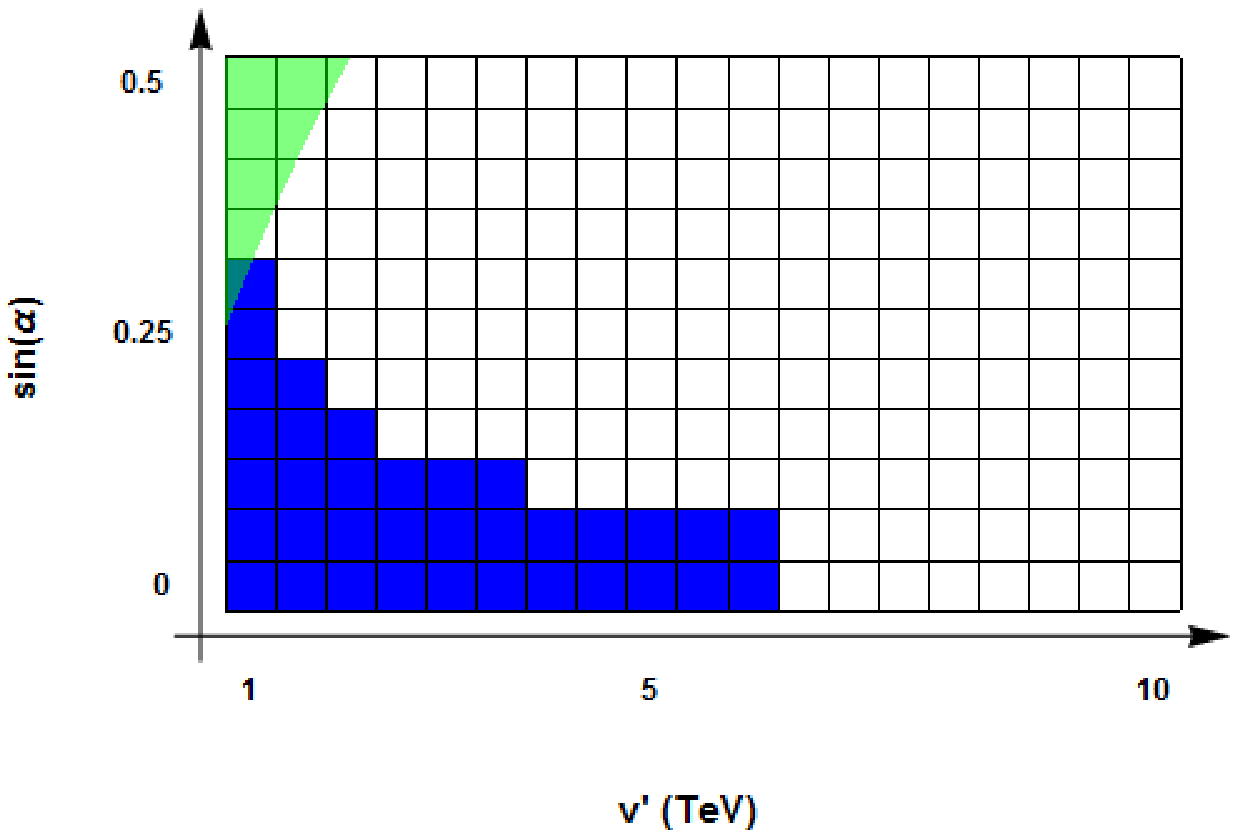}
   \caption{The matrix {plots} for showing the region where a proper amount of the relic density is realized when $m_H = 250\,\text{GeV},\, m_{k^{\pm\pm}} = 500\,\text{GeV}$.
Conventions are the same with ones in Fig.~\ref{fig:DM_matrixplot_MH250GeV_Mk250GeV}.
{No excluded region is found by the direct detection in the shown parameter range.}
}
   \label{fig:DM_matrixplot_MH250GeV_Mk500GeV}
\end{figure}

\begin{figure}[t]
\centering
\includegraphics[width=0.495\columnwidth]{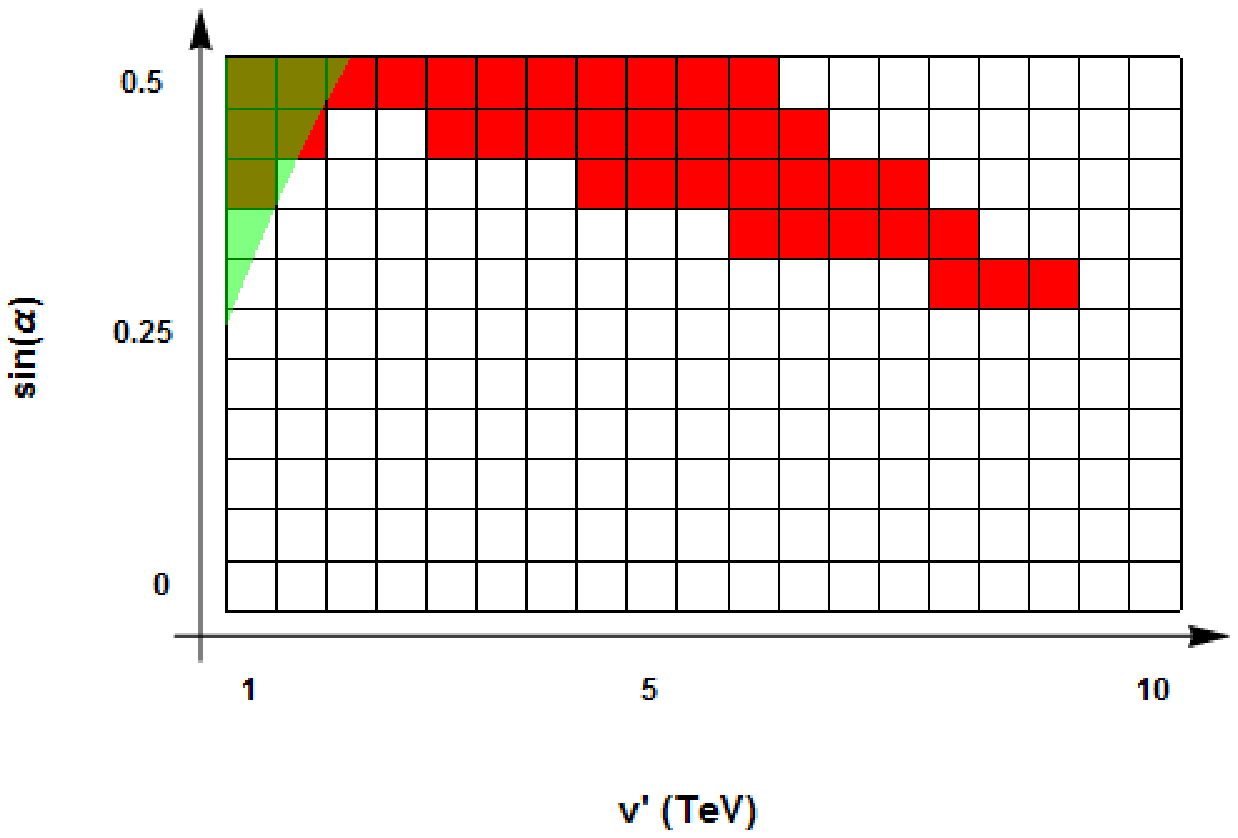}
\includegraphics[width=0.495\columnwidth]{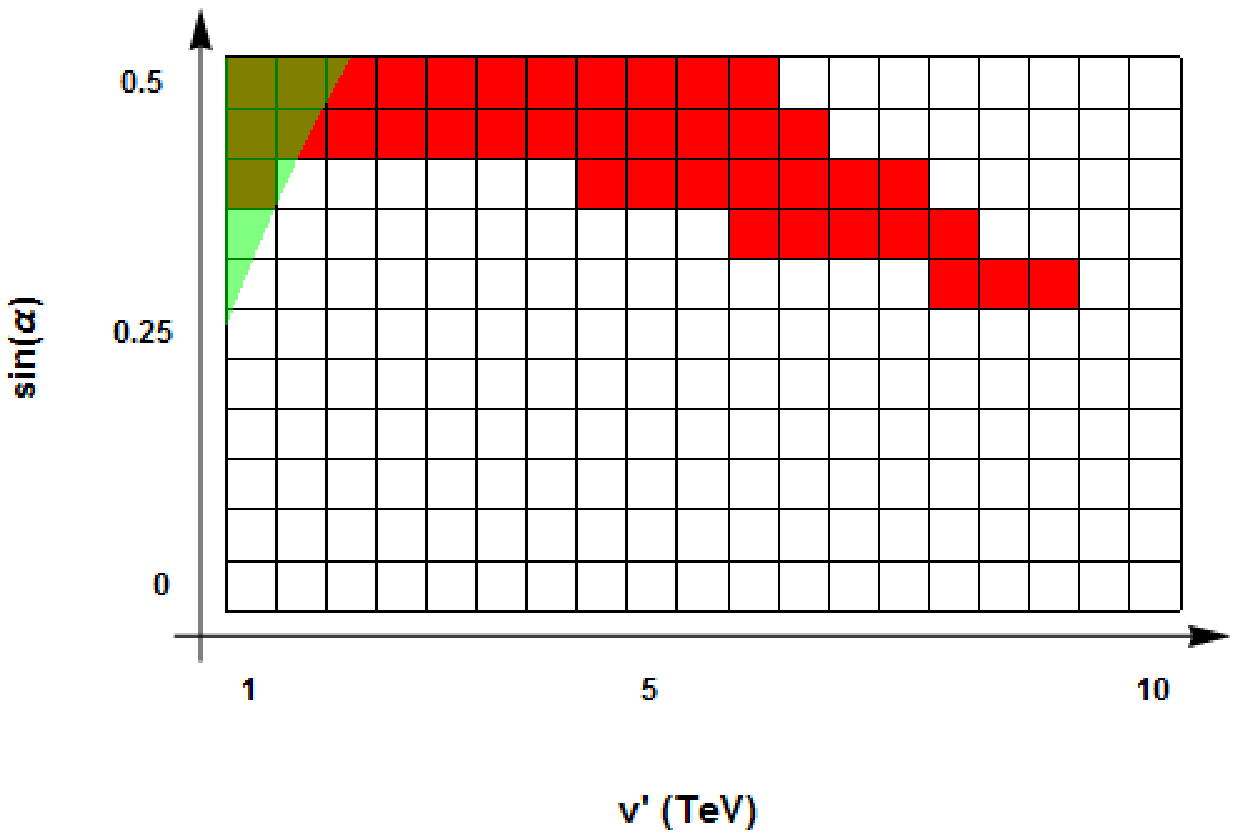}
   \caption{Similar plots as Figs.~\ref{fig:DM_matrixplot_MH250GeV_Mk250GeV} and \ref{fig:DM_matrixplot_MH250GeV_Mk500GeV} when $m_H = 500\,\text{GeV},\, m_{k^{\pm\pm}} = 250\,\text{GeV}$ (left) and $m_H = 500\,\text{GeV},\, m_{k^{\pm\pm}} = 500\,\text{GeV}$ (right), respectively.
Here, no solution is found around $m_\chi \simeq m_H/2 \simeq 250\,\text{GeV}$, while one around $m_\chi \simeq m_h/2 \simeq 62.5\,\text{GeV}$ is still there.
The green region is excluded by the invisible decay of the $125\,\text{GeV}$ Higgs.
{No excluded region is found by the direct detection in the shown parameter range.}
}
   \label{fig:DM_matrixplot_MH500GeV}
\end{figure}

In the following calculation, we treat the variables $\alpha$, $m_H${,} and $v'$ independently, which {determines} the coefficients $\lambda_{\Phi\Sigma}$, $\lambda_\Phi${,} and $\lambda_\Sigma$ through the relations in Eqs.~(\ref{eq:CP-even_matrix}){--}(\ref{eq:CP-even_mixing}) as
\begin{align}
\lambda_{\Phi\Sigma} &= \frac{\sin{\alpha} \cos{\alpha} \, (m_H^2 - m_h^2)}{v v'}, \notag \\
\lambda_{\Phi} &= \frac{\cos^2{\alpha} \, m_h^2 + \sin^2{\alpha} \, m_H^2}{2v^2}, \notag \\
\lambda_{\Sigma} &= \frac{\sin^2{\alpha} \, m_h^2 + \cos^2{\alpha} \, m_H^2}{2{v'}^2}.
\end{align}
{We set two quartic couplings $\lambda_{\Phi k}$ and $\lambda_{\Sigma k}$ as $1.0$ and $0.1$.}
In our scenario, $v'$ tends to be $\mathcal{O}(1)\,\text{TeV}$ leading to {suppression of} the {thermally averaged} cross section describing pair annihilation processes{;} see Appendix~\ref{sec:appendix_amplitude} for detail forms.
It would require a mechanism for enhancing the cross section.
{Hence, in} this manuscript, we focus on the two resonant regions of $m_\chi$ around $m_h/2$ or $m_H/2$.
Our requirement for the amount of the relics is that it should be within the $\pm2\sigma$ range of the latest value reported by the Planck experiment {[$0.1196 \pm 0.0031\,(68\%\,\text{C.L.})$~\cite{Ade:2013zuv}]}.
In the following matrix plots, the values of $v'$ and $\sin{\alpha}$ at the center of a square represent the inputs in the whole region shown by the square.

An important point is that all the effective diagrams of this {pair annihilation} of $\chi$ are {Higgs mediated}.\footnote{{Note that the processes $N_{R_1} \bar{N}_{R_1} \to \ell^{+}_{i} \ell^{-}_{j}\,(i,j=1,2,3)$ with exchanging $h^{\pm}_2$ in $t$-channel, which are important in similar situations~\cite{Krauss:2002px,Cheung:2004xm}, are ineffective and neglected in the present scenario because the lower bound on $m_{h^\pm_2}$ is around a few TeV and decoupled as shown in subsection~\ref{sec:scanning}.}}
Like the couplings between the observed fermions and the $125\,\text{GeV}$ Higgs {boson} in {the} SM, the DM-portal scalar couplings are proportional to the factor $m_\chi/v'$.
In Fig.~\ref{fig:reliccurve}, we present realized present-day relic densities of the dark matter candidate $\chi$ as functions of $m_\chi$, where the other effective parameters are fixed as $\sin{\alpha} = 0.4,\, v' = 800\,\text{GeV},\, m_H = 250\,\text{GeV},\, m_{k^{\pm\pm}} = 500\,\text{GeV}$ (left panel) and $\sin{\alpha} = 0.5,\, v' = 5\,\text{TeV},\, m_H = 250\,\text{GeV},\, m_{k^{\pm\pm}} = 500\,\text{GeV}$ (right panel), respectively.
Here, we can see that a suppression due to a huge $v'$ in the latter case and the point around the heavier resonance of $H$ is not suitable for a correct amount of DM today.
Around the lighter one of $h$, less $s$-channel propagator suppression could be expected and thus the solution around this pole {still} remains.

Another critical variable in the relic calculation is the mixing angle $\alpha$ among the two {$CP$ even} scalars $h$ and $H$.
If $\sin{\alpha}$ is close to zero, contributions from the mediator $h$ are diminished significantly due to almost zeroness of the couplings between $\chi$ and the SM particles in the final state.
Thereby, a suitable amount of the mixing is requested for the solution near $m_h/2$.
In the following analysis, we focus on non-negative values in $\sin\alpha$ since the sign of $\sin\alpha$ is insignificant in the resonant regions.\footnote{{We mention details of this point. The sign of $\sin\alpha$ affects results only through the amplitude of $\chi \bar{\chi} \to h h$ and the partial decay width of $h/H \to \gamma \gamma$ and $h/H \to Z \gamma$ in the $s$-channel propagators.
In the DM masses which we consider, the amplitude gives only a subleading contribution.
The effect coming from the widths is also negligible since the branching ratios of $h/H \to \gamma \gamma$ and $h/H \to Z \gamma$ are pretty small.}
}
Also, we ignore the range of $|\sin\alpha| > 0.5$ since no allowed region was found in the results of the global analyses shown in Figs.~\ref{fig:LHCHiggscontour_1}{--}\ref{fig:LHCHiggscontour_2}.

As actual examples, we investigate the four possibilities in $(m_H, m_{k^{\pm\pm}})$ of $(250\,\text{GeV}, 250\,\text{GeV})$ in Fig.~\ref{fig:DM_matrixplot_MH250GeV_Mk250GeV}; $(250\,\text{GeV}, 500\,\text{GeV})$ in Fig.~\ref{fig:DM_matrixplot_MH250GeV_Mk500GeV}; {and} $(500\,\text{GeV}, 250\,\text{GeV})$ and $(500\,\text{GeV}, 500\,\text{GeV})$ in Fig.~\ref{fig:DM_matrixplot_MH500GeV}, respectively.
The regions covered with {the color green} are excluded, where the branching ratio of the observed Higgs {boson} to invisible pairs {exceeds} the $95\%$ confidence level upper bound $0.29$ reported by the ATLAS experiment~\cite{ATLAS:2015yda}.\footnote{In this estimation, the Standard Model production cross section was assumed in $m_h = 125\,\text{GeV}$. Here, we do not care about this point seriously and simply utilize this result for putting a bound on the present scenario.}
The value of $v'$ {and $\alpha$} almost completely determines the profiles since the primary final state with invisibleness is a pair of the NG {bosons} $G$.
{We mention} that no exclusion from the direct detection process is found in the shown parameter ranges since the elastic scattering cross section is suppressed by the large VEV $v'$.
We observe small differences originating from the value of $m_{k^{\pm\pm}}$ since in the DM mass range of $m_\chi \lesssim m_{k^{\pm\pm}}$, $k^{\pm\pm}$ can contribute to the relic density {only} indirectly through the total widths $\Gamma_{h}$ and $\Gamma_{H}$.\footnote{We simply ignore the process $\chi\chi \to k^{++} k^{--}$ in the case near the kinetic boundary, $m_H = 500\,\text{GeV}$ and $m_{k^{\pm\pm}} = 250\,\text{GeV}$, where the heavier resonant region corresponds to $m_\chi \simeq 250\,\text{GeV}$.
{Here, we} do not consider the lighter $k^{\pm\pm}$ with $100 \sim 250\,\text{GeV}$ mass, {the possibility of which was examined in the global fits in section~\ref{sec:LHCglobalfit}.}
}

A {suggestive} aspect via the results is that only around the $m_h/2$ is possible when $m_H$ gets large as $500\,\text{GeV}$ because the suppression via {the} heavy resonant particle could eliminate the solution near $m_H/2$.
On the other hand, as discussed in {Sec.}~\ref{sec:LHCglobalfit}, the $7$ and $8\,\text{TeV}$ LHC results {restrict} the scalar mixing angle $\alpha$, where we quote a typical bound we obtained, $|\sin{\alpha}| \lesssim 0.3$.
For making the {DM} solution around $m_h/2$ realizable, we should select {the absolute value of the mixing angle $|\sin{\alpha}|$} around or more than $0.3$ as shown in Figs.~\ref{fig:DM_matrixplot_MH250GeV_Mk250GeV}{--}\ref{fig:DM_matrixplot_MH500GeV}.
{Let us remind {the reader} that the sign of $\sin\alpha$ does not affect the result (calculated in $\sin\alpha \geq 0$) significantly.}
In addition, here, the value of $v'$ is also constrained as around or less than $9\,\text{TeV}$ in the case of $m_\chi \simeq m_h/2$.
In conclusion, the present scenario is still viable, but a part of the parameter space generating a suitable amount of relics {comes} {to be} on the edge.
Note that our calculation would {be} applicable for a rather general setup of a fermion dark matter communicating the SM sector through two scalars with renormalizable interactions and $\mathcal{O}(1)$ TeV $v'$ after the decoupling of $k^{\pm\pm}$.

\section{Summary and Conclusion \label{sec:conclusion}}

In this paper, we proposed a {three loop} radiative neutrino mass scenario with an {\it isolated} {doubly charged} singlet scalar $k^{\pm\pm}$ without couplings to the charged leptons at the leading order for solving 
an undesirable situation in our previous study~\cite{Hatanaka:2014tba}.
The previous scenario has the same particle contents of charged singlet scalars with the present model, two {singly charged} ones ($h_1^\pm$, $h_2^\pm$) and one {doubly charged} one ($k^{\pm\pm}$), but assignments of additional global symmetries are different, which makes a difference in interactions associated with the charged bosons.

In the previous case without Majorana neutrinos, $k^{\pm\pm}$ should attach the charged leptons at the leading order to generate a neutrino mass matrix at the {three loop} level.
But a side effect due to this type of couplings with LFV could be serious from a phenomenological point of view.
{Because of} resultant tree-level contributions to LFV processes, related couplings are severely constrained, where realized neutrino masses get to be smaller than the observed values.
{Increasing} scalar trilinear couplings with charged scalars appearing in the numerator of elements of the loop-induced neutrino mass matrix seems to be an option.
However, the largeness in the trilinear couplings tends to destabilize the vacuum rapidly.
Making all of the charged scalars very heavy is a solution since the scalar trilinear couplings do not contribute to LFV at the leading order and the ratio of {the trilinear couplings divided by the masses} of the charged scalars can remain ``sizable" for the observed neutrino masses.
Here, however, a typical mass scale of the charged bosons should be around $10\,\text{TeV}$, where we cannot expect a signal suggesting the existence of the particles at {the} LHC.

Our modified setup in the assignment of a global $U(1)$ symmetry brings us a better situation.
Introducing Majorana neutrinos allows $k^{\pm\pm}$ to be away from the charged leptons.
{Now} {the LFV is relatively relaxed,}
and at least a part of {the} couplings contributing to the neutrino mass can take larger values.
Thus, scalar trilinear couplings need not to be drastically large, and{,} as a result, vacuum destabilization is alleviated.
We {found}
that a few hundred GeV $k^{\pm\pm}$ is realizable, which can be tested at the ongoing LHC experiment, even though the two {singly charged} scalars are still very heavy {at} around a few TeV.
{It is noteworthy that the associated NG boson $G$ does not generate dangerous interactions {that} severely restrict the decay constant $v'$.
The constant $v'$ can takes a value around the TeV scale with no harm.
}

Another constraint on the system comes from the mixing of two {$CP$} even neutral scalars, which are mixed states {between}
the component of the $SU(2)_L$ doublet $\Phi$ in {the} SM and the corresponding part of a complex $SU(2)_L$ singlet scalar $\Sigma_0$ for {the} spontaneous symmetry breaking of the global $U(1)$.
After the breakdown, Majorana fermions obtain masses via the Higgs mechanism and a residual discrete parity ensures the stability of an accidental DM candidate of the lightest Majorana neutrino $N_{R_1}$.
This mixing angle plays a very important role in the DM-related phenomena in this model since the candidate can couple to the SM sector only through the two mass eigenstates of the {$CP$} even scalars $h\,(=125\,\text{GeV})$ and $H$.
{Note that the two {singly charged} scalars should be decoupled in our scenario and diagrams associated with them are negligible.}

The latest LHC result of the Higgs search by the ATLAS and the CMS experiments {puts} a stringent bound on the mixing angle $\alpha$, $|\sin{\alpha}| \, \lesssim \, 0.3$ as a roughly estimated bound through a global analysis. 
A typical scale of {the} VEV of $\Sigma_0$ would be around a few TeV scale, where the DM pair annihilation cross section is significantly suppressed since the {Yukawa} couplings between $N_{R_1}$ and $\Sigma_0$ are inversely proportional to {the} VEV.
When the DM mass is around the two resonant regions, $m_h/2$ and $m_H/2$, the cross section is enhanced very much.
But we found 
that when $M_{H}$ is heavy {at} around $500\,\text{GeV}$, the possibility near $m_H/2$ is closed due to the propagator suppression of the massive mediator.
On the other hand, {to activate} the solution around $m_h/2$, the mixing angle $\alpha$ should be large to a certain degree {to ensure} a sizable interaction between the DM and {SM particles}.
Typically, a required value in this direction is $|\sin{\alpha}| \, \gtrsim \, 0.3$, which would not have large overlaps with the preferred area obtained via the global analysis on the Higgs search results.

In spite of {the} insignificance of the bounds via the invisible $125\,\text{GeV}$ Higgs decay and  the DM direct detection because of the coupling suppression via the huge VEV of $\Sigma_0$, experimental data in Higgs physics put sizable constraints on our scenario.
In the near future, updated Higgs results could declare validity of our scenario much more precisely.
Also, searching for a suitable way to discriminate the present scenario from other models with charged scalars 
will be an important task at the LHC and other future colliders.

%
%

\vspace{0.3cm}
\section*{Acknowledgments}

K.\,N. is grateful {for} Tomohiro Abe, Motoi Endo, {Masahiro Ibe}, Shinya Kanemura, Tetsuo Shindou, Hiroaki Sugiyama, Koji Tsumura and Toshifumi Yamashita for kind suggestions and useful discussions.
H.\,O. {is sincerely grateful for} all the KIAS members, Korean cordial persons, foods, culture, weather, and all the other things.
This work is supported in part by NRF Research No. 2009-0083526 (Y.\,O.) of the Republic of Korea.

\appendix
\section{Analytic forms of loop functions for LFV \label{sec:appendix_loopfunction}}

In this appendix, we show analytic forms of the loop functions {for processes with LFV}.
The functions for $\ell \to \ell \gamma$ in Eq.~(\ref{eq:ltolgamma_constraint}) {are} obtained as
\begin{align}
I'_{1,a} &= {\frac{1}{(4\pi)^2}} {\int_{0}^{1} \!\! dx \int_{0}^{1-x} \!\!\!\!\!\!\! dy}\  \frac{x(2x-1)}{(x+y) m_{h_2^{\pm}}^2 + (1-x-y) M_{N_a}^2} \notag \\
&=
\begin{cases}
{\frac{1}{(4\pi)^2}}
\frac{9 m_{h_2^\pm}^4 M_{N_a}^2
   + 9 m_{h_2^\pm}^2 M_{N_a}^4 \left(2 \log \left( \frac{M_{N_a}^2}{m_{h_2^\pm}^2} \right) +1 \right)
   + M_{N_a}^6 \left(6 \log \left( \frac{M_{N_a}^2}{m_{h_2^\pm}^2} \right) -17 \right)
   - m_{h_2^\pm}^6}{36 \left(m_{h_2^\pm}^2 - M_{N_a}^2 \right)^4} & (\text{for } m_{h_2^\pm} \not= M_{N_a}), \\
0 & (\text{for } m_{h_2^\pm} = M_{N_a}),
\end{cases} \\
I'_{2,a} &= {\frac{1}{(4\pi)^2}} {\int_{0}^{1} \!\! dx \int_{0}^{1-x} \!\!\!\!\!\!\! dy}\  \frac{x(2y-1)}{(x+y) m_{h_2^{\pm}}^2 + (1-x-y) M_{N_a}^2} \notag \\
&=
\begin{cases}
{\frac{1}{(4\pi)^2}}
\frac{27 m_{h_2^\pm}^4 M_{N_a}^2
   - 9 m_{h_2^\pm}^2 M_{N_a}^4 \left(2 \log \left( \frac{m_{h_2^\pm}^2}{M_{N_a}^2} \right) +3 \right)
   + M_{N_a}^6 \left(6 \log \left( \frac{m_{h_2^\pm}^2}{M_{N_a}^2} \right) +5 \right)
   - 5 m_{h_2^\pm}^6}{36 \left(m_{h_2^\pm}^2 - M_{N_a}^2 \right)^4} & (\text{for } m_{h_2^\pm} \not= M_{N_a}), \\
- {\frac{1}{(4\pi)^2}} \frac{1}{12 m_{h_2^\pm}^2} & (\text{for } m_{h_2^\pm} = M_{N_a}).
\end{cases}
\end{align}
Note that when we take the limit $M_{N_a} \to 0$, $I'_{1,a}$ and $I'_{2,a}$ are reduced {to} $I_{1,a}$ and $I_{2,a}$ (up to the difference of $m_{h_1^\pm}$ and $m_{h_2^\pm}$), respectively,
\begin{align}
I'_{1,a} \to  - {\frac{1}{(4\pi)^2}} \frac{1}{36 m_{h_2^\pm}^2},\quad
I'_{2,a} \to  - {\frac{1}{(4\pi)^2}} \frac{5}{36 m_{h_2^\pm}^2}.
\end{align}

The functions $J_{1,ab}$ and $J_{2,ab}$ for describing {the $\ell \to 3\ell$} processes in Eq.~(\ref{eq:lto3l_constraint}) take the following forms
\begin{align}
J_{1,ab} &=
\frac{1}{(4\pi)^2} {\int_{0}^{1} \!\! dx \int_{0}^{1-x} \!\!\!\!\!\!\! dy} \  \frac{1-x-y}{\left[x M_{N_{a}}^2 + y M_{N_{b}}^2 + (1-x-y) m_{h_2^{\pm}}^2\right]} \notag \\
&=
{\scriptstyle
{
\frac{1}{(4\pi)^2} \bigg\{
\frac{m_{h_2^\pm}^6 M_{N_a}^2 + m_{h_2}^4 M_{N_a}^4 \left( \log\left(\frac{M_{N_a}^2}{m_{h_2^\pm}^2}\right) -1 \right) - m_{h_2^\pm}^6 M_{N_b}^2 + m_{h_2^\pm}^2 M_{N_a}^4 M_{N_b}^2 \left( 2\log\left(\frac{m_{h_2^\pm}^2}{M_{N_a}^2}\right) +1 \right)}
{2 \left( M_{N_a}^2 - M_{N_b}^2 \right) \left( m_{h_2^\pm}^2 - M_{N_a}^2 \right)^2 \left( m_{h_2^\pm}^2 - M_{N_b}^2 \right)^2} }} \notag \\
&\qquad
{\scriptstyle
{
+
\frac{m_{h_2^\pm}^4 M_{N_b}^4 \left( \log\left(\frac{m_{h_2^\pm}^2}{M_{N_b}^2}\right) +1 \right) + m_{h_2^\pm}^2 M_{N_a}^2 M_{N_b}^4 \left( 2\log\left(\frac{M_{N_b}^2}{m_{h_2^\pm}}\right) -1 \right) + M_{N_a}^4 M_{N_b}^4 \log\left(\frac{M_{N_a}^2}{M_{N_b}^2}\right) }
{2 \left( M_{N_a}^2 - M_{N_b}^2 \right) \left( m_{h_2^\pm}^2 - M_{N_a}^2 \right)^2 \left( m_{h_2^\pm}^2 - M_{N_b}^2 \right)^2} \bigg\}
}}
\notag \\
& \hspace{92mm} (\text{for } m_{h_2^\pm} \not= M_{N_a} \not= M_{N_b}),
\end{align}
\begin{align}
& {J_{2,ab}} =
{\frac{1}{(4\pi)^2}} {\int_{0}^{1} \!\! dx \int_{0}^{1-x} \!\!\!\!\!\!\! dy} \  \frac{1-x-y}{\left[x M_{N_{a}}^2 + y M_{N_{b}}^2 + (1-x-y) m_{h_2^{\pm}}^2\right]^2} \notag \\
& \hspace{-23mm} =
{\scriptstyle
\frac{m_{h_2^\pm}^4 \left(M_{N_a}^2 \left( \log \left(\frac{m_{h_2^\pm}^2}{M_{N_a}^2} \right) -1 \right)
   +M_{N_b}^2 \left( \log \left( \frac{M_{N_b}^2}{m_{h_2^\pm}^2} \right) +1 \right) \right)
   +m_{h_2^\pm}^2 \left(2 M_{N_a}^2 M_{N_b}^2
   \log \left( \frac{M_{N_a}^2}{M_{N_b}^2} \right)+M_{N_a}^4-M_{N_b}^4\right)
   +M_{N_a}^2 M_{N_b}^4 \left( \log \left( \frac{m_{h_2^\pm}^2}{M_{N_a}^2} \right) +1 \right)
   +M_{N_a}^4 M_{N_b}^2 \left( \log \left( \frac{M_{N_b}^2}{m_{h_2^\pm}^2} \right) -1 \right)}
   {{(4\pi)^2}
   \left(m_{h_2^\pm}^2-M_{N_a}^2\right)^2
   \left(m_{h_2^\pm}^2-M_{N_b}^2\right)^2
   \left(M_{N_a}^2-M_{N_b}^2\right)}
} \notag \\
& \hspace{100mm} (\text{for } m_{h_2^\pm} \not= M_{N_a} \not= M_{N_b}),
\end{align}
where the following specific cases are also obtained:
\begin{align}
J_{1,ab} &\to
\begin{cases}
\frac{1}{(4\pi)^2}
\frac{2 m_{h_2^\pm}^2 M_{N_a}^2 \log
   \left(\frac{M_{N_a}^2}{m_{h_2^\pm}^2}\right)-M_{N_a}^4+m_{h_2^\pm}^4}{2
   \left(m_{h_2^\pm}^2-M_{N_a}^2\right)^3}
   & (\text{for } M_{N_a} = M_{N_b} \not= m_{h_2^\pm}), \\
\frac{1}{(4\pi)^2}
\frac{1}{2 m_{h_2^\pm}^2} & (\text{for } M_{N_a} = M_{N_b} = m_{h_2^\pm}),
\end{cases} \\
{J_{2,ab}} &\to
\begin{cases}
{\frac{1}{(4\pi)^2}}
\frac{m_{h_2^\pm}^2 \left(\log \left( \frac{m_{h_2^\pm}^2}{M_{N_a}^2} \right) -2 \right)
   +M_{N_a}^2 \left( \log \left( \frac{m_{h_2^\pm}^2}{M_{N_a}^2} \right) +2 \right)}
   {\left(m_{h_2^\pm}^2-M_{N_a}^2\right)^3}
   & (\text{for } M_{N_a} = M_{N_b} \not= m_{h_2^\pm}), \\
{\frac{1}{(4\pi)^2}}
\frac{1}{6 m_{h_2^\pm}^4} & (\text{for } M_{N_a} = M_{N_b} = m_{h_2^\pm}).
\end{cases}
\end{align}

\section{Partial widths of CP even scalars \label{sec:appendix_width}}

In this appendix, we represent a part of partial decay widths with nontrivial forms of the two {$CP$ even} scalars $h$ and $H$ with our notation for loop functions.
\begin{align}
\Gamma_{h \to GG}
&=
\frac{m_h^3}{{32} \pi v'^2} \sin^2{\alpha}, \\
\Gamma_{h \to N_{R_1} \bar{N}_{R_1}}
&=
\frac{1}{{16}\pi} \left( \frac{M_{N_1}}{v'} \right)^2 m_h \left( 1 - \frac{4 M_{N_{1}}^2}{m_h^2} \right)^{3/2} \sin^2{\alpha}, \\
\Gamma_{H \to GG}
&=
\frac{m_H^3}{{32} \pi v'^2} \cos^2{\alpha}, \\
\Gamma_{H \to N_{R_1} \bar{N}_{R_1}}
&=
\frac{1}{{16} \pi} \left( \frac{M_{N_1}}{v'} \right)^2 m_H \left( 1 - \frac{4 M_{N_{1}}^2}{m_H^2} \right)^{3/2} \cos^2{\alpha}, \\
\Gamma_{H \to 2h}
&=
\frac{1}{{32} \pi m_H} |C_{Hhh}|^2 \sqrt{1 - \frac{4m_h^2}{m_H^2}}, \\
\Gamma_{h \to gg}
&=
\frac{G_{F} \alpha_{s}^2 m_h^3}{36\sqrt{2}\pi^3} \left|\frac{3}{4} \left( A_{1/2}^{\gamma\gamma}(\tau_t) \right) \cos{\alpha} \right|^2, \\
\Gamma_{h \to \gamma\gamma}
&=
\frac{G_{F} \alpha_{\text{EM}}^2 m_h^3}{128\sqrt{2}\pi^3} \left|\left( A_1^{\gamma\gamma}(\tau_W) + N_C Q_t^2 A_{1/2}^{\gamma\gamma}(\tau_t) \right) \cos{\alpha} + \frac{1}{2} \frac{v^2}{m_{k^{\pm\pm}}^2} Q_k^2 C_{hkk} A_0^{\gamma\gamma}(\tau_k)\right|^{2}, \\
\Gamma_{h \to Z\gamma}
&=
\frac{{\alpha_{\text{EM}}^2} m_{h}^3}{512\pi^3} \left( 1 - \frac{m_Z^2}{m_h^2} \right)^3
\left| {A_{\text{SM}}^{Z\gamma}} \cos{\alpha} - \frac{v}{m_{k^{\pm\pm}}^2} \left( 2 Q_k g_{Zkk} \right) C_{hkk} A_0^{Z\gamma}(\tau_k,\lambda_k) \right|^2, \\
\Gamma_{H \to gg}
&=
\frac{G_{F} \alpha_{s}^2 m_H^3}{36\sqrt{2}\pi^3} \left|\frac{3}{4} \left( A_{1/2}^{\gamma\gamma}(\tau'_t) \right) \sin{\alpha} \right|^2, \\
\Gamma_{H \to \gamma\gamma}
&=
\frac{G_{F} \alpha_{\text{EM}}^2 m_H^3}{128\sqrt{2}\pi^3} \left|\left( A_1^{\gamma\gamma}(\tau'_W) + N_C Q_t^2 A_{1/2}^{\gamma\gamma}(\tau'_t) \right) \sin{\alpha} + \frac{1}{2} \frac{v^2}{m_{k^{\pm\pm}}^2} Q_k^2 C_{Hkk} A_0^{\gamma\gamma}(\tau'_k)\right|^{2}, \\
\Gamma_{H \to Z\gamma}
&=
\frac{{\alpha_{\text{EM}}^2} m_{H}^3}{512\pi^3} \left( 1 - \frac{m_Z^2}{m_H^2} \right)^3
\left| {A_{\text{SM}}^{Z\gamma}}(\tau_{W,t} \to \tau'_{W,t})  \sin{\alpha} - \frac{v}{m_{k^{\pm\pm}}^2} \left( 2 Q_k g_{Zkk} \right) C_{Hkk} A_0^{Z\gamma}(\tau'_k,\lambda_k) \right|^2,
\end{align}
with $C_{hkk}$ in Eq.~(\ref{effective_couplings_oneloopHiggs}) and the parameters
\begin{align}
\tau'_{i} = \frac{4m_i^2}{m_{H}^2}\ (i=t,W,k),\quad
C_{Hkk} = \sin{\alpha} \lambda_{\Phi k} + \cos{\alpha} \lambda_{\Sigma k} \left(\frac{v'}{v}\right).
\end{align}
{$G_F$ and $\alpha_s$ are the Fermi constant and the QCD coupling strength, respectively.}
Note that in the calculation $\Gamma_{H \to Z\gamma}$, the replacement $\tau_i \to \tau'_i$ in the factor $A_{\text{SM}}^{Z\gamma}$ is required.
One can refer to {Sec.}~\ref{sec:LHCglobalfit} for the detail of loop functions.

\section{Matrix elements for $N_{R_1}$ dark matter annihilations \label{sec:appendix_amplitude}}

In this appendix, we summarize the averaged matrix elements squared for relic density calculation of the dark matter candidate $N_{R_1}$ through pair annihilations.
Note that no symmetric factor is included originating from identical final states {in the following forms}, which should be considered in the integration over the solid angle for estimating the thermal average of the cross section $\langle \sigma v_{\text{rel}} \rangle$.
\begin{align}
|\overline{\mathcal{M}}(N_{R_1} \bar{N}_{R_1} \to f\bar{f})|^2
&=
4 {N_{C_f}} \left( \frac{M_{N_{1}} m_f s_{\alpha} c_{\alpha}}{vv'} \right)^2
\left| \frac{1}{s - m_h^2 + im_h\Gamma_h} - \frac{1}{s - m_H^2 + im_H\Gamma_H} \right|^2 \times \notag \\
&\quad \left[ (p_1 \cdot p_2) - M_{N_{1}}^2 \right] \left[ (k_1 \cdot k_2) - m_f^2 \right], \\
|\overline{\mathcal{M}}(N_{R_1} \bar{N}_{R_1} \to ZZ)|^2
&=
\left( \frac{M_{N_{1}} g_2 M_W s_{\alpha} c_{\alpha}}{v' c_W^2} \right)^2
\left| \frac{1}{s - m_h^2 + im_h\Gamma_h} - \frac{1}{s - m_H^2 + im_H\Gamma_H} \right|^2 \times \notag \\
&\quad \left[ (p_1 \cdot p_2) - M_{N_{1}}^2 \right] \left[ 2 + \frac{(k_1 \cdot k_2)^2}{M_Z^4} \right], \\
|\overline{\mathcal{M}}(N_{R_1} \bar{N}_{R_1} \to W^+W^-)|^2
&=
\left( \frac{M_{N_{1}} g_2 M_W s_{\alpha} c_{\alpha}}{v'} \right)^2
\left| \frac{1}{s - m_h^2 + im_h\Gamma_h} - \frac{1}{s - m_H^2 + im_H\Gamma_H} \right|^2 \times \notag \\
&\quad \left[ (p_1 \cdot p_2) - M_{N_{1}}^2 \right] \left[ 2 + \frac{(k_1 \cdot k_2)^2}{M_W^4} \right], \\
|\overline{\mathcal{M}}(N_{R_1} \bar{N}_{R_1} \to GG)|^2
&=
4 \left( \frac{M_{N_{1}}}{vv'} \right)^2 (k_1 \cdot k_2)^2
\left| \frac{s_\alpha^2}{s - m_h^2 + im_h\Gamma_h} + \frac{c_\alpha^2}{s - m_H^2 + im_H\Gamma_H} \right|^2 \times \notag \\
&\quad \left[ (p_1 \cdot p_2) - M_{N_{1}}^2 \right], \\
|\overline{\mathcal{M}}(N_{R_1} \bar{N}_{R_1} \to hh)|^2
&=
\left( \frac{M_{N_{1}}}{v'} \right)^2
\left| \frac{s_\alpha C_{hhh}}{s - m_h^2 + im_h\Gamma_h} - \frac{c_\alpha C_{Hhh} }{s - m_H^2 + im_H\Gamma_H} \right|^2 \times \notag \\
&\quad \left[ (p_1 \cdot p_2) - M_{N_{1}}^2 \right],
\end{align}
where $N_{C_f}$ shows the color factor $3$ ($1$) when $f$ is a quark (lepton), $p_1, p_2$ and $k_1, k_2$ are {initial- and final-state} momenta, respectively.
$s$ is {defined} as $s = (p_1 + p_2)^2 = (k_1 + k_2)^2$.
Here, we use the {shorthand} notations of $c_\alpha \equiv \cos{\alpha}$ and $s_{\alpha} \equiv \sin{\alpha}$.
The forms of effective couplings describing scalar self interactions are
\begin{align}
C_{hhh}
&=
3 \lambda_{\Phi \Sigma}
\left( v (s_\alpha^2 c_\alpha) - v' (c_{\alpha}^2 s_\alpha) \right)
{+ 6\lambda_{\Phi} v c_\alpha^3 - 6\lambda_{\Sigma} v' s_\alpha^3}, \\
C_{Hhh}
&=
\lambda_{\Phi \Sigma} {\left( v \left( s^3_{\alpha} - 2 c^2_{\alpha} s_{\alpha} \right) +
v' \left( c_{\alpha}^3 -2 c_{\alpha} s^2_{\alpha} \right) \right)
+ 6\lambda_{\Phi} v c_\alpha^2 s_{\alpha} + 6\lambda_{\Sigma} v' s_\alpha^2 c_\alpha}.
\end{align}
{We mention that the processes $N_{R_1} \bar{N}_{R_1}$ into $H H$, $h H$, $k^{++} k^{--}$, $h_1^+ h_1^-$ and $h_2^+ h_2^-$ are ineffective in our discussion when the mass of the DM is around $m_h/2$ or $m_H/2$  considering the bounds discussed in {Sec.}~\ref{sec:neutrino_fitting}{--}\ref{sec:LHCglobalfit}.}

Note that the processes $N_{R_1} \bar{N}_{R_1} \to \ell^{+}_{i} \ell^{-}_{j}\,(i,j=1,2,3)$ with {the exchange of} $h^{\pm}_2$ in {the $t$ channel}, which are important in similar situations~\cite{Krauss:2002px,Cheung:2004xm}, are ineffective and neglected in the present scenario because the lower bound on $m_{h^\pm_2}$ is around a few TeV and decoupled as shown in {Sec.}~\ref{sec:scanning}.
We neglect the three- and four-body final states via virtual $W$ and $Z$ boson decays, which gives sizable modifications near the thresholds for producing gauge boson pairs~\cite{Cline:2012hg,Cline:2013gha}, since our interest is only around $m_h/2$ and $m_H/2$ where this correction is expected to be {ineffective}.


\end{document}